\documentclass[12pt,journal,onecolumn,draftclsnofoot]{IEEEtran}
\usepackage[noadjust]{cite}
\usepackage{cite}
\usepackage{amsfonts}
\usepackage{amssymb}
\usepackage{eurosym}
\usepackage{graphicx}
\usepackage{epstopdf}
\usepackage{amsmath}
\usepackage{tikz,lipsum}
\usepackage[labelformat=simple]{subcaption}
\usepackage{float}
\usepackage[T1]{fontenc}
\usepackage{amsthm}
\usepackage{mathrsfs}
\usepackage{mathrsfs}
\usepackage{lipsum}
\usepackage{ragged2e}
\usepackage{mathtools}
\usepackage{multicol}
\usepackage{listings}
\usepackage{algorithm,algorithmic}
\usepackage{authblk}
\usepackage{times}
\usepackage{latexsym}
\usepackage{array}

\bstctlcite{IEEEexample:BSTcontrol}
\ifCLASSOPTIONcompsoc
\else
\fi
\ifCLASSOPTIONcompsoc
\else
\fi
\ifCLASSINFOpdf
\else
\fi
\newtheorem{theorem}{Theorem}

\newtheorem{remark}{Remark}
\newcounter{mytempeqncnt}
\begin{document}

\title{A PHY Layer Security Analysis of Uplink Cooperative Jamming-Based
Underlay CRNs with Multi-Eavesdroppers}
\author{ Mounia~Bouabdellah, Faissal~El~Bouanani,~\IEEEmembership{Senior
Member,~IEEE,}~and~Mohamed-Slim~Alouini,~\IEEEmembership{Fellow,~IEEE} 
\IEEEcompsocitemizethanks{\IEEEcompsocthanksitem M. Bouabdellah and F. El Bouanani are with ENSIAS College of Engineering, Mohammed V University, Rabat, Morocco (e-mails: mounia\_bouabdellah@um5.ac.ma, f.elbouanani@um5s.net.ma).\protect \and
		\IEEEcompsocthanksitem M.-S. Alouini is with Computer, Electrical, and Mathematical Sciences and Engineering (CEMSE) Division, King Abdullah University of Science and Technology (KAUST), Thuwal 23955-6900, Makkah Province, Saudi Arabia (e-mail: slim.alouini@kaust.edu.sa).
}}
\maketitle

\begin{abstract}
In this paper, the physical layer security of a dual-hop underlay uplink
cognitive radio network is investigated over Nakagami-$m$ fading channels.
Specifically, multiple secondary sources $(S_{i})_{1\leq i \leq N}$ are
taking turns in accessing the licensed spectrum of the primary users and
communicating with a multi-antenna secondary base station ($D$) through the
aid of {a multi-antenna} relay $R$ in the presence of %
{$M$} eavesdroppers 
{$(E_{k})_{1\leq k \leq
M}$} {that are also equipped with multiple antennas}. Among
the remaining nodes, one jammer is randomly selected to transmit an
artificial noise to disrupt all the eavesdroppers that are attempting to
intercept the communication of the legitimate links i.e., $S_{i}$-$R$ and $R$%
-$D$. The received signals at {each node} are combined using
maximum-ratio combining. 
{Secrecy analysis is provided by
deriving} closed-form and asymptotic expressions for the secrecy outage
probability. The impact of several key parameters on the system's secrecy
e.g., transmit power of the sources, number of eavesdroppers, maximum
tolerated interference power, and the number of diversity branches is
investigated. 
{Importantly, by considering two scenarios,
namely (i) absence and (ii) presence of a friendly jammer, new insights are
obtained for the considered communication system. Especially, we tend to answer to the following question: Can better secrecy be achieved without jamming by considering a single antenna at {eavesdroppers} and multiple-ones at the legitimate users (i.e., relay and end-user) rather than sending permanently an artificial noise and considering that both the relay and the destination are equipped with a single antenna, while multiple antennas are used by the eavesdroppers?}
The obtained results are corroborated through Monte Carlo simulation %
{and} show that the system's security {can be}
enhanced by adjusting the aforementioned parameters.
\end{abstract}

\begin{IEEEkeywords}
	Cognitive Radio Networks, Eavesdropping, Jamming signals, Physical Layer Security, Secrecy outage probability.
\end{IEEEkeywords}

\section{Introduction}

The increasing number of mobile users led to an unprecedented demand for
spectral resources. In this regard, cognitive radio has emerged as a new
paradigm that enhances the spectrum efficiency by allowing its reuse \cite%
{mitola2002cognitive}. In underlay cognitive radio networks (CRNs), the
issue of radio-frequency spectrum scarcity is alleviated by allowing the
secondary users (SUs) to share the spectrum with primary users (PUs) under
the condition of not causing any harmful interference to them. Consequently,
the SUs are required to continuously adjust their transmit powers in order
to meet the PUs' quality of service (QoS). Under such constraints, ensuring
the physical layer security (PLS) of multi-hop CRNs becomes a challenge of
utmost importance. To remedy this problem, several techniques can be used to
strengthen the secrecy capacity at each hop namely increasing the number of
diversity branches at the receivers, sending a jamming signal with the
highest power, increasing maximum transmit power at the source and maximum
tolerated interference power as well, reducing the number of hops, employing
zero-forcing precoding techniques, involving energy harvesting (EH) and
non-orthogonal multiple access (NOMA) technique etc.

Recently, the PLS of CRNs has been the focus of many recent research works.
For instance, non-cooperative CRNs were considered in \cite{lei2016secrecy}-%
\cite{nguyen2017secure}, therein all receivers i.e., both destination and
eavesdropper were assumed to be equipped with multiple antennas and perform
the selection combining (SC) technique. Particularly, in \cite%
{lei2016secrecy}, the source is also assumed to be a multi-antennas node
performing transmit antenna selection, while in \cite{tran2017cognitive} the
secrecy performance is investigated for both secondary and primary networks.
Closed-form and asymptotic expressions for the secrecy outage probability
(SOP) were derived under Rayleigh \cite{tran2017cognitive}, \cite%
{elkashlan2014security} and Nakagami-$m$ \cite{lei2016secrecy}, \cite%
{nguyen2017secure} fading models.

The PLS of multi-relays dual-hop CRNs was explored in \cite{lei2017secrecy}-%
\cite{sakran2012proposed}. Specifically, in \cite{lei2017secrecy} and \cite%
{ho2017analysis}, the communication was performed in the presence of only
one eavesdropper attempting to overhear the communication channel, while
multiple eavesdroppers were considered in \cite{ding2016secrecy} and \cite%
{sakran2012proposed}. Furthermore, In \cite{lei2017secrecy}, optimal and
suboptimal relay selection were analyzed while in \cite{ho2017analysis} the
relay that minimizes the signal-to-noise ratio (SNR) of the wiretap link was
chosen. Besides, in \cite{ding2016secrecy}, the most threatening
eavesdropper is selected first according to the maximum SNR of the wiretap
links between the source and the eavesdroppers. Next, the best relay
minimizing the SNR at the selected eavesdropper is then chosen. In \cite%
{sakran2012proposed}, the relay that maximizes the achievable secrecy rate
is selected. Under these conditions, closed-form and asymptotic expressions
for the SOP and {intercept probability (IP)} were derived
over either Nakagami-$m$ \cite{lei2017secrecy} or Rayleigh \cite%
{ho2017analysis}-\cite{sakran2012proposed} fading channels. The IP and SOP
analysis of cooperative underlay EH-based CRNs have been investigated in 
\cite{TVT2} and \cite{TVT2}-\cite{access}, respectively. Specifically, the
SUs have been assumed to harvest energy from the PU's signals in \cite{TVT2}-%
\cite{tgcn}. In contrast, in \cite{ISWCS}-\cite{access} the relay is
harvesting energy from the SU signals instead.

{The PLS of NOMA-based CRNs has been investigated in
\cite{Xiang2019}-\cite{Song2019}. In \cite{Xiang2019}, an overlay NOMA CRN
was considered such that the SUs were assumed to be eavesdroppers, while the
PLS of mmWave NOMA CRN was investigated in \cite{Song2019}. Closed-form
expressions for the connection outage probability, SOP and secrecy throughput
were derived over Nakagami-$m$ fading channels. }

PLS analysis through the aid of a friendly jammer was discussed in \cite%
{zou2016physical}-\cite{MouniaBouabdellah}. In \cite{zou2016physical}, the
IP was derived by considering multiple source-destination pairs
communicating under eavesdropping attempts of only one eavesdropper, with
the source cooperation aided opportunistic jamming. In \cite{liu2014relay},
the SOP of dual-hop aided opportunistic jamming CRNs is investigated. In
this work, one relay is selected to forward the information while another
one is chosen to disrupt the eavesdropper by sending an artificial noise.
Also, in the two aforementioned works, several selection policies of the
friendly jammer were considered. The impact of the friendly jammer's
transmit power in the presence of multiple eavesdroppers by considering a
direct communication link between multiple sources and one destination is
discussed in \cite{MouniaBouabdellah}.

In this work, we investigate the joint impact of the friendly jammer's
transmit power, multiple SUs with power adaptation constraint, number of
eavesdroppers, number of diversity branches, maximum tolerated interference
power at the PU receiver on the PLS of a cooperative underlay uplink CRNs
under Nakagami-$m $ fading model. Without loss of generality, it is worth
mentioning that 
{each user is transmitting its data independently from other
users}. Consequently, the sources are assumed to transmit in turns their
data while a friendly jammer is randomly selected among the remaining idle
sources to transmit an artificial noise so as to disrupt the eavesdroppers. 
{In this scheme, the nodes $ R $, $D$, $E_{k} $ perform MRC
technique, hence the knowledge of the channel state information (CSI) at
these nodes is necessary. For this reason, we assume that the CSI is
available. Additionally, eavesdroppers {are considered are
passive}}. 

The main contributions of this paper can be summarized as follows:

\begin{itemize}
\item The PLS of an underlay uplink dual-hop CRN operating under Nakagami-$m$
fading environment is investigated by deriving closed-form and asymptotic
expressions for the SOP of the overall system under two scenarios namely,
(i) presence and (ii) absence of a friendly jammer.

\item Under the power adaptation constraint of the SUs, the joint impact of
the discussed parameters on the system's security is investigated.

\item We show that the system's security is enhanced in the presence of an
important number of eavesdroppers by increasing the (i) SUs' transmit powers
(ii) number of legitimate destination branches (iii) and maximum tolerated
interference power.
\end{itemize}

The rest of this paper is organized as follows. In Section II, the system
and channel models are presented. Closed-form as well as asymptotic
expressions for the SOP are derived in Section III. In Section IV, the
numerical and simulation results are provided and discussed for various key
parameters' values. Finally, this work is concluded in Section V.

\section{System and channel models}

The considered two-hops CRN, represented in Fig. 1, consists of multiple
sources $(S_{i})_{i=1,..,N}$, one {$L_{R}$-antennas} relay $R$%
, multiple {$L_{E_{k}}$-antennas} eavesdroppers $%
(E_{k})_{k=1,..,M}$, one destination $D$ equipped with $L_{D}$ antennas,
one PU transmitter $(P_{Tx})$, and one PU receiver $(P_{Rx})$.%
{ For the sake of simplicity, we assume that the relay receives the transmitted signals from $ S_{i} $ on the $ L_{R} $ antennas and uses only one antenna
to forward the message to $ D $}. Moreover, we consider multi-user
scheduling such that, at any given moment, only one user is transmitting its
data. Also, the source nodes are taking rounds in accessing the spectrum and
a friendly jammer $S_{J}$ is randomly selected among $N-1$ remaining nodes
to send an artificial noise. This latter can be canceled by legitimate
nodes, while $E_{k}$ cannot mitigate it, leading to an increase %
{in} the secrecy capacity. 
{Similarly to \cite{zou2016physical}, we
assume that a friendly jammer generates an artificial noise using a
pseudo-random sequence that is known to the legitimate users which allows
them to cancel out this noise, while this sequence remains unknown to the
illegitimate ones. To this end, the main aim} of this work is to investigate
the impact of 
{a friendly jammer, legitimate, and wiretap
channels' average SNRs, maximum tolerated interference power as well as the
spatial diversity at both the relay and the end-user} on the secrecy
performance of the considered communication system. In this scheme, Nakagami-%
$m$ fading model is considered for all links. The fading amplitudes of links 
$S_{i}\rightarrow R$, $R\rightarrow (D_{t})_{1\leq t\leq L_{D}}$, $%
S_{i}\rightarrow E_{k}$, $R\rightarrow E_{k}$, $R\rightarrow P_{Rx}$, $%
S_{i}\rightarrow P_{Rx}$ are denoted by $h_{q}$ where $q=\{S_{i}R,$ $RD_{t},$
$S_{i}E_{k},$ $RE_{k},$ $RP, $ $S_{i}P\}$. Consequently, the channel gains $%
g_{q}=\left\vert h_{q}\right\vert ^{2}$ are Gamma distributed with
probability density function (PDF) and cumulative density function (CDF) are
given by 
\begin{equation}
f_{g_{q}}(x)=\frac{\lambda _{q}^{m_{q}}}{\Gamma (m_{q})}x^{m_{q}-1}e^{-%
\lambda _{q}x},  \label{f_gq}
\end{equation}%
\begin{equation}
F_{g_{q}}\left( x\right) =\frac{\gamma \left( m_{q},\lambda _{q}x\right) }{%
\Gamma \left( m_{q}\right) },  \label{F_gq}
\end{equation}%
where $\lambda _{q}=\dfrac{m_{q}}{\Omega _{q}}$, $m_{q}$ and $\Omega _{q}$
denote the fading severity and the average channel power gain, respectively, 
$\Gamma \left( .\right) $ and $\gamma \left( .,.\right) $ are the Euler and
the lower incomplete Gamma functions \cite[Eqs. (8.310.1), (8.350.1)]{table}%
, respectively. For a natural number $m_{q},$ the above CDF can be written
as [13, Eq. (8.352.1)] 
\begin{equation}
F_{g_{q}}\left( x\right) =1-e^{-\lambda _{q}x}\sum_{k=0}^{m_{q}-1}\frac{%
\lambda _{q}^{k}x^{k}}{k!}.  \label{F_gq_exp2}
\end{equation}%
The received signals at $R$, $E_{k}$ at both hops, and $D$ are given,
respectively, by 
\begin{equation}
y_{R}^{(i)}=\sqrt{P_{S_{i}}}{\mathbf{||h}_{S_{i}R}\mathbf{||}}%
x_{S_{i}}+{\mathbf{w}_{S_{i}R}}\mathbf{n}_{R},\text{ }i=1,..,N{,}
\label{Y_R}
\end{equation}%
\begin{align}
y_{1E_{k}}^{(i)}& =\sqrt{P_{S_{i}}}{\mathbf{||h}_{S_{i}E_{k}}%
\mathbf{||}}x_{S_{i}}+\epsilon \sqrt{P_{S_{J}}}{%
\mathbf{||h}_{S_{J}E_{k}}\mathbf{||}}x_{S_{J}}+{%
\mathbf{w}_{S_{i}E_{k}}}\mathbf{n}_{E_{k}}, \\
& k=1,..,M,\text{ }i=1,..,N,\text{ }J\neq i{,}  \notag  \label{3}
\end{align}%
\begin{equation}
y_{2E_{k}}=\sqrt{P_{R}}{\mathbf{||h}_{RE_{k}}\mathbf{||}}%
x_{R}+{\mathbf{w}_{RE_{k}}}\mathbf{n}_{E_{k}},k=1,..,M{,}
\label{Y_2Ek}
\end{equation}%
\begin{equation}
y_{D}=\sqrt{P_{R}}\mathbf{||h}_{RD}\mathbf{||}x_{R}+\mathbf{w}_{RD}\mathbf{n}%
_{D},  \label{Y_D}
\end{equation}%
with 
\begin{equation*}
\epsilon =%
\begin{cases}
0,\text{ Absence of a jammer} \\ 
1,\text{ Presence of a jammer}%
\end{cases}%
{.}
\end{equation*}%
Here, $P_{n}$ and $x_{n}$ denote the transmit power and signal from the node 
$n$, respectively where $n=\{S_{i},R\}$, 
{$\mathbf{w}_{q}=\frac{\mathbf{h}_{q}^{\dagger }}{||\mathbf{h}_{q}||}$, $q=\{S_{i}R,S_{i}E_{k},RE_{k},RD\},$
while $\mathbf{h}_{q}$ denotes $L_{n}\times 1,$ channel vector of
the links $S_{i}$-$R$, $S_{i}$-$E_{k}$, $R$-$D$}, $\dagger $ denotes the
transpose conjugate, and $||.||$ represents the Frobenius norm. Also, $%
\mathbf{n}_{R}$, $\mathbf{n}_{D}$, and $\mathbf{n}_{E_{k}}$, denote the %
{$ N_{n}\times1 $} additive white Gaussian noise %
{vector} at $R$, $D$, and $E_{k}$, respectively. 
{ For the sake of simplicity, all noise power vectors'
components are considered equal $N_{0}$}. 

Throughout the transmission process, both $S_{i}$, and $R$ have to adapt
their transmit powers so as to avoid causing harmful interference to the
PUs. Thus, the transmit power of the source and the relay $R$ taking into
consideration the maximum constraint power can be, respectively, expressed as

\begin{equation}
P_{S_{i}}{}=\min \left( P_{S_{i}}^{max},\,\dfrac{P_{I}}{g_{S_{i}P}}\right)
;i=1,..,N,  \label{P_Si}
\end{equation}%
and 
\begin{equation}
P_{R}=\min \left( P_{R}^{max},\,\dfrac{P_{I}}{g_{RP}}\right) ,  \label{P_R}
\end{equation}%
where $P_{S_{i}}^{max}$ and $P_{R}^{max}$ denote the maximum transmit power
at $S_{i}$, and $R$, respectively, while $P_{I}$ accounts for the maximum
tolerated interference power at $P_{Rx}$. 
\begin{figure}[tbh]
\begin{centering}
		\includegraphics[scale=0.3]{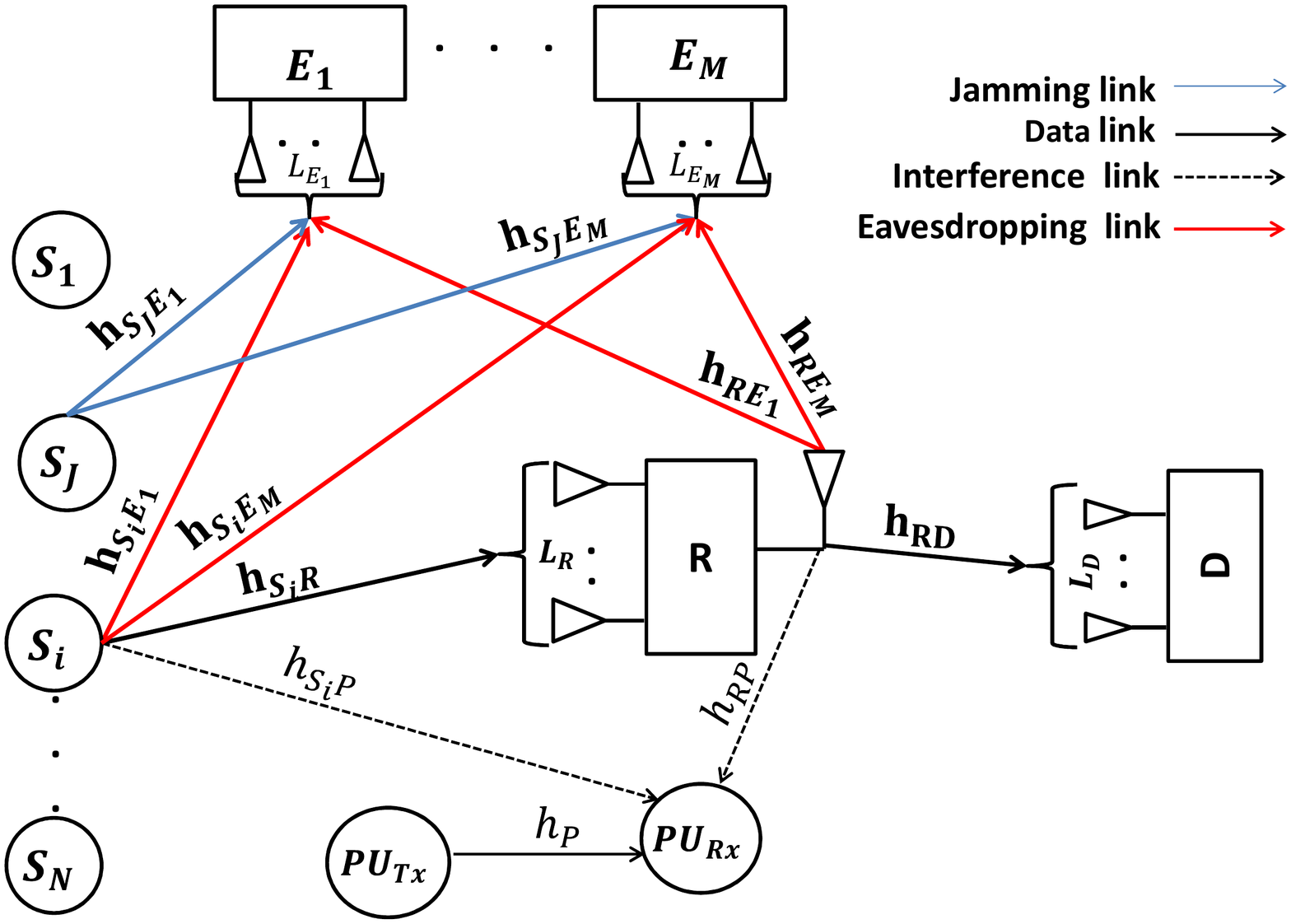}
		\par\end{centering}
\centering{}
\caption{{System setup.}}
\end{figure}

\section{Secrecy outage probability}

In this section, we start by defining the secrecy capacities for the two
hops' links. Next, we present the SOP analysis in terms of both closed-form
and asymptotic expressions for various cases of fading severity parameters,
specifically, those of the legitimate and the wiretap links of the first
hop. Also, two scenarios will be considered, namely, the presence and
absence of a friendly jammer.

\subsection{Secrecy Capacity}

The secrecy capacity can be defined as the maximum rate at which the
transmitted information can confidentially reach its intended destination.
In our considered system, the secrecy capacities in the case of presence and
absence of a friendly jammer are given, respectively, by 
\begin{equation}
C_{s}^{(i,J)}=\underset{k=1,..,M}{\min }\left(
C_{1S}^{(i,\,k,J)},\,C_{2S}^{(k)}\right) ,  \label{Cs_i}
\end{equation}
\begin{equation}
C_{s}^{(i)}=\underset{k=1,..,M}{\min }\left(
C_{1S}^{(i,\,k)},\,C_{2S}^{(k)}\right) ,  \label{Cs_nojammer}
\end{equation}
where

\begin{itemize}
\item $C_{1S}^{(i,\,k,J)}$ and $C_{1S}^{(i,\,k)}$ denote the secrecy
capacities at the first hop, i.e., the difference between the capacity of
the main link $S_{i}-R$ and the one of the wiretap channel $S_{i}-E_{k}$ in
the presence and absence of a friendly jammer, respectively, and can be
written as 
\begin{equation}
C_{1S}^{(i,k,J)}=%
\begin{cases}
\log _{{\small 2}}\left( \dfrac{1+\gamma _{R}^{(i)}}{1+\gamma _{1E}^{(i,k,J)}%
}\right) {\small ,}\text{ }{\small \gamma }_{R}^{(i)}{\small >\gamma }%
_{1E}^{(i,k,J)} \\ 
0,\text{ elsewhere}%
\end{cases}%
,
\end{equation}%
\begin{equation}
C_{1S}^{(i,k)}=%
\begin{cases}
\log _{{\small 2}}\left( \dfrac{1+\gamma _{R}^{(i)}}{1+\gamma _{1E}^{(i,k)}}%
\right) {\small ,}\text{ }{\small \gamma }_{R}^{(i)}{\small >\gamma }%
_{1E}^{(i,k)} \\ 
0,\text{ elsewhere}%
\end{cases}%
,
\end{equation}%
where $\gamma _{R}^{(i)}$ denotes the instantaneous SNR at $R$, while $%
\gamma _{1E}^{(i,k,J)}$ and $\gamma _{1E}^{(i,k)}$ stand for the SNRs at the
eavesdropper $E_{k}$ in the presence and absence of a friendly jammer,
respectively, and are given by 
\begin{equation}
{\gamma _{R}^{(i)}=\min \left( \overline{{\small \gamma
}}_{S_{i}},\frac{\overline{{\small \gamma }}_{I}}{g_{S_{i}P}}\right)
\sum_{u=1}^{L_{R}}g_{S_{i}R_{u}}},  \label{Gamma_R}
\end{equation}%
\begin{equation}
{ \gamma_{1E}^{(i,k,J)}=\dfrac{\min \left( \overline{{\small
\gamma }}_{S_{i}},\frac{\overline{{\small \gamma }}_{I}}{g_{S_{i}P}}\right)
\sum_{u=1}^{L_{E_{k}}}g_{S_{i}E_{k}^{(u)}}}{\min \left( \overline{{\small
\gamma }}_{S_{J}},\frac{\overline{{\small \gamma }}_{I}}{g_{S_{J}P}}\right)
\sum_{u=1}^{L_{E_{k}}}g_{S_{J}E_{k}^{(u)}}+1},}  \label{Gamma_E1}
\end{equation}
\begin{equation}
{ \gamma _{1E}^{(i,k)}=\min \left( \overline{{\small \gamma
}}_{S_{i}},\frac{ \overline{{\small \gamma }}_{I}}{g_{S_{i}P}}\right)
\sum_{u=1}^{L_{E_{k}}}g_{S_{i}E_{k}^{(u)}},}  \label{Gamma_ik}
\end{equation}
and $\overline{{\small \gamma }}_{S_{i}}=P_{S_{i}}^{max}/N_{0}$, $\overline{ 
{\small \gamma }}_{I}=P_{I}/N_{0},$ and $\overline{{\small \gamma }}
_{S_{J}}=P_{S_{J}}^{max}/N_{0}.$ 

\item $C_{2S}^{(k)}$ is the secrecy capacity of the second hop, representing
the difference between the capacity of the link $R-D$ and the one of the
wiretap channel $R-E_{k}$ 
\begin{equation}
{\small C}_{2S}^{(k)}{\small =}%
\begin{cases}
\log _{2}\left( \dfrac{1+\gamma _{D}}{1+\gamma _{2E}^{(k)}}\right) {\small ,}%
\text{ }{\small \gamma }_{D}>\gamma _{2E}^{(k)} \\ 
0,\text{ elsewhere}%
\end{cases}%
{\small ,}  \label{C_2S_R_k}
\end{equation}%
where $\gamma _{D},$ and $\gamma _{2E}^{(k)}$ denote the instantaneous SNR
of the main link $R-D$ and the channel $R-E_{k}$, respectively and are given
as 
\begin{equation}
\gamma _{D}=\min \left( \overline{\gamma }_{R},\frac{\overline{\gamma }_{I}}{%
g_{RP}}\right) \sum_{t=1}^{L_{D}}g_{RD_{t}},  \label{Gamma_D}
\end{equation}%
\begin{equation}
{\gamma_{2E}^{(k)}=\min
\left(\overline{\gamma}_{R},\frac{\overline{\gamma}_{I}}{g_{RP}}\right)
\sum_{u=1}^{L_{E_{k}}}g_{RE_{k}^{(u)}},}  \label{gamma2E}
\end{equation}%
with $\overline{\gamma }_{R}=P_{R}^{\max }/N_{0}.$
\end{itemize}

\begin{remark}
\begin{itemize}
\item One can see from (\ref{Gamma_R}) and (\ref{Gamma_E1}), that the PHY
layer security at the first hop in the presence of a friendly jammer can be
enhanced by increasing separately $\overline{\gamma }_{I}$, $\overline{%
{\small \gamma }}_{S_{i}},$ or $\overline{{\small \gamma }}_{S_{J}}.$
Indeed, the increasing scale of the SNR at the relay exceeds the one of the $%
k$th eavesdropper as a jamming signal is added to the one received by $%
E_{k}. $ However, in the absence of a friendly jammer, one can see from (\ref%
{Gamma_R}) and (\ref{Gamma_ik}) that only the impact of legitimate and
wiretap channels' parameters can make the distinction between the two
associated SNRs. Consequently, the smaller $\lambda _{S_{i}R}$, the greater
the secrecy capacity and then the security gets improved.

\item From (\ref{Gamma_D}) and (\ref{gamma2E}), it can be noticed that
increasing either $\overline{\gamma }_{R}$ or $\overline{\gamma }_{I}$
enhances more the capacity of the legitimate link as $D$ performs the MRC
technique. Additionally, increasing the number of antennas at the receiver
increases the SNR at $D $. Consequently, the system's security gets enhanced
as well.
\end{itemize}
\end{remark}

\subsection{Exact Secrecy Outage Probability}

In this paper, the SOP is chosen as a performance metric and it accounts for
the probability that the secrecy capacity is less than a predefined secrecy
rate $R_{s}.$ For the considered system, the $N$ sources are taking rounds
in accessing the spectrum then one jammer is randomly selected among the $%
N-1 $ remaining sources. The SOP if there were no jamming, can be expressed
as 
\begin{equation}
SOP=\frac{1}{N}\sum_{i=1}^{N}SOP^{(i)},
\end{equation}%
while in the presence of a jammer, it becomes \cite{zou2016physical} 
\begin{equation}
SOP=\frac{1}{N(N-1)}\sum_{i=1}^{N}\sum_{\substack{ J=1  \\ J\neq i}}%
^{N}SOP^{(i,J)},  \label{SOP_i_expp1}
\end{equation}%
where $SOP^{(i)}$ and $SOP^{(i,J)}$ account for SOP of the system linking $%
S_{i}$ with $D$ in the presence of eavesdroppers, and in the absence and
presence of the $J$th friendly jammer, respectively. The SOP of the
considered system stands for the probability that at least one of the
secrecy capacities falls below a predefined secrecy rate $R_{s},$ namely 
\begin{align}
SOP^{(i,J)}& =1-\prod_{k=1}^{M}\Pr \left( \min
(C_{1S}^{(i,\,k,J)},\,C_{2S}^{(k)})\geq R_{s}\right)  \label{SOPikJ} \\
& =1-\prod_{k=1}^{M}\left[ 1-SOP_{1}^{(i,k,J)}\right] \left[ 1-SOP_{2}^{(k)}%
\right] ,  \notag
\end{align}%
and 
\begin{equation}
SOP^{(i)}=1-\prod_{k=1}^{M}\left[ 1-SOP_{1}^{(i,k)}\right] \left[
1-SOP_{2}^{(k)}\right] ,  \label{SOP_i}
\end{equation}%
where $SOP_{1}^{(i,k,J)}$ and $SOP_{1}^{(i,k)}$ stand for the secrecy
capacities at the first hop in the presence and absence of a friendly
jammer, respectively, and $SOP_{2}^{(k)}$ represents the secrecy capacity at
the second hop. One can see from (\ref{SOPikJ}) and (\ref{SOP_i}) that the
computation of SOP requires the knowledge of $SOP_{1}^{(i,k,J)},$ $%
SOP_{1}^{(i,k)},$ and $SOP_{2}^{(k)}$ as well.

\begin{remark}
As $SOP_{1}^{(i,k,J)},$ $SOP_{1}^{(i)},$ and $SOP_{2}^{(k)}$ are between $0 $
and $1 $, it is worth mentioning that the greater is $M $, the greater is
SOP (approaches 1), and then the system becomes vulnerable to eavesdropping
attack.
\end{remark}

\begin{theorem}
The closed-form expressions of $SOP_{1}^{(i,k,J)},$ $SOP_{1}^{(i,k)},$ and $%
SOP_{2}^{(k)}$ under Nakagami-$m$ fading model are given by (\ref%
{SOP_1k_final}), (\ref{SOP1_ik_final}), and (\ref{SOP2_final}),
respectively, as shown at the top of the next page, where $\varpi
_{i}^{(k)}=\gamma \lambda _{S_{i}R}+\lambda _{S_{i}E_{k}},$ $\sigma _{i}=%
\overline{{\small \gamma }}_{I}/\overline{{\small \gamma }}_{S_{i}},$ $%
\delta =\overline{{\small \gamma }}_{I}/\overline{{\small \gamma }}_{R},$ $%
\gamma =2^{R_{S}},$ $\theta _{i}^{(k,J)}=\lambda _{S_{J}E_{k}}/\left( 
\overline{{\small \gamma }}_{S_{J}}\lambda _{S_{i}E_{k}}\right) ,$ $\varphi
_{J}=\lambda_{S_{J}P}\overline{{\small \gamma }}_{I}/\overline{{\small %
\gamma }}_{S_{J}},\varsigma _{i}^{(k,J)}=\lambda _{S_{J}E_{k}}/\left(
\lambda_{S_{J}P}\lambda _{S_{i}E_{k}}\right) ,$ $\varphi _{R}=\lambda _{RP}%
\overline{{\small \gamma }}_{I}/\overline{{\small \gamma }}_{R}$, $\varphi
_{S_{i}}=\lambda _{S_{i}P}\overline{\gamma }_{P}/\overline{\gamma }_{S_{i}},$
\end{theorem}

\begin{figure*}[t]
\setcounter{mytempeqncnt}{\value{equation}} \setcounter{equation}{23}
\par
\begin{eqnarray}
SOP_{1}^{(i,k,J)} &=&1-\frac{\gamma \left( m_{S_{i}P},\varphi
_{S_{i}}\right) \Gamma \left( {L_{R}m_{S_{i}R}},\frac{\sigma
_{i}\xi _{S_{i}R}}{\overline{{\small \gamma }}_{I}}\right) +\mathcal{M}%
_{3}\left( \frac{\xi _{S_{i}R}}{\lambda _{S_{i}P}\overline{\gamma }_{I}}%
\right) }{\Gamma ({L_{R}m_{S_{i}R}})\Gamma \left(
m_{S_{i}P}\right) }+\frac{\gamma \alpha _{i}^{(k,J)}}{\Gamma \left(
m_{S_{i}P}\right) }\sum_{h{\small =0}}^{%
{L_{E_{k}}m_{S_{i}E_{k}}}{\small -1}}\Omega
_{h}^{(i,k)}\sum_{l=0}^{{L_{R}m_{S_{i}R}}-1}\frac{\Upsilon
_{l}^{(i)}\left( \overline{{\small \gamma }}_{I}\right) ^{-%
{L_{R}m_{S_{i}R}}+l+1}}{\left( \varpi _{i}^{(k)}\right) ^{%
{L_{E_{k}}m_{S_{i}E_{k}}}+l-h}}  \notag \\
&&\times \left[ \frac{\gamma \left( m_{S_{i}P},\varphi _{S_{i}}\right) e^{-%
\frac{\sigma _{i}\xi _{S_{i}R}}{\overline{{\small \gamma }}_{I}}}}{\sigma
_{i}^{-{L_{R}m_{S_{i}R}}+l+1}}+\frac{\lambda
_{S_{i}P}^{m_{S_{i}P}}\Gamma \left( m_{S_{i}P}+%
{L_{R}m_{S_{i}R}}-l-1,\varphi _{S_{i}}+\frac{\sigma _{i}\xi
_{S_{i}R}}{\overline{\gamma }_{I}}\right) }{\left( \lambda _{S_{i}P}+\frac{%
\xi _{S_{i}R}}{\overline{\gamma }_{I}}\right) ^{m_{S_{i}P}+%
{L_{R}m_{S_{i}R}}-l-1}}\right]  \notag \\
&&\times \left[ \gamma \left( m_{S_{J}P},\varphi _{J}\right) \mathcal{M}%
_{1}^{(h,l)}\left( \varpi _{i}^{(k)}\theta _{i}^{(k,J)}\right) +\mathcal{M}%
_{2}^{(h,l)}\left( \frac{\varsigma _{i}^{(k,J)}\varpi _{i}^{(k)}}{\overline{%
{\small \gamma }}_{I}}\right) \right] .  \label{SOP_1k_final}
\end{eqnarray}%
\par
\hrulefill{} \vspace*{4pt}
\end{figure*}
\begin{figure*}[t]
\setcounter{mytempeqncnt}{\value{equation}} \setcounter{equation}{24}
\par
\begin{eqnarray}
SOP_{1}^{(i,k)} &=&1-\frac{\lambda _{S_{i}R}^{%
{L_{R}m_{S_{i}R}}}\gamma }{\Gamma \left( {%
{L_{E_{k}}m_{S_{i}E_{k}}}}\right) \Gamma (%
{L_{R}m_{S_{i}R}})\Gamma \left( m_{S_{i}P}\right) }%
\sum_{l=0}^{{L_{R}m_{S_{i}R}}-1}\frac{\Upsilon _{l}^{(i)}%
\overline{{\small \gamma }}_{I}^{-{L_{R}m_{S_{i}R}}+l+1}}{%
\left( \gamma \lambda _{S_{i}R}\right) ^{l+1}}G_{2,2}^{1,2}\left( \frac{%
\lambda _{S_{i}E_{k}}}{\lambda _{S_{i}R}\gamma }\left\vert 
\begin{array}{c}
-l,1;- \\ 
{{L_{E_{k}}m_{S_{i}E_{k}}}};0%
\end{array}%
\right. \right)  \label{SOP1_ik_final} \\
&&\times \left[ \frac{\gamma \left( m_{S_{i}P},\varphi _{S_{i}}\right) e^{-%
\frac{\sigma _{i}\xi _{S_{i}R}}{\overline{{\small \gamma }}_{I}}}}{\sigma
_{i}^{-{L_{R}m_{S_{i}R}}+l+1}}+\frac{\lambda
_{S_{i}P}^{m_{S_{i}P}}\Gamma \left( m_{S_{i}P}+%
{L_{R}m_{S_{i}R}}-l-1,\varphi _{S_{i}}+\frac{\sigma _{i}\xi
_{S_{i}R}}{\overline{{\small \gamma }}_{I}}\right) }{\left( \lambda
_{S_{i}P}+\frac{\xi _{S_{i}R}}{\overline{{\small \gamma }}_{I}}\right)
^{m_{S_{i}P}+{L_{R}m_{S_{i}R}}-l-1}}\right] .  \notag
\end{eqnarray}%
\par
\hrulefill {} \vspace*{4pt}
\end{figure*}
\begin{figure*}[t]
\setcounter{mytempeqncnt}{\value{equation}} \setcounter{equation}{25}
\par
\begin{eqnarray}
SOP_{2}^{(k)} &=&1-\frac{\lambda _{RD}^{L_{D}m_{RD}}}{\Gamma \left( %
{L_{E_{k}}m_{RE_{k}}}\right) \Gamma (L_{D}m_{RD})\Gamma
(m_{RP})}\sum_{j=0}^{L_{D}m_{RD}-1}\frac{\mathcal{B}_{j}\overline{{\small %
\gamma }}_{I}^{-L_{D}m_{RD}+j+1}}{\lambda _{RD}^{j+1}}G_{2,2}^{1,2}\left( 
\frac{\lambda _{_{RE_{k}}}}{\lambda _{RD}\gamma }\left\vert 
\begin{array}{c}
-j,1;- \\ 
{L_{E_{k}}m_{RE_{k}}};0%
\end{array}%
\right. \right)  \label{SOP2_final} \\
&&\times \left[ \frac{e^{-\frac{\delta \xi _{RD}}{\overline{{\small \gamma }}%
_{I}}}\gamma \left( m_{RP},\varphi _{R}\right) }{\delta ^{-L_{D}m_{RD}+j+1}}+%
\frac{\lambda _{RP}^{m_{RP}}\Gamma \left( L_{D}m_{RD}+m_{RP}-j-1,\varphi
_{R}+\frac{\delta \xi _{RD}}{\overline{{\small \gamma }}_{I}}\right) }{%
\left( \frac{\xi _{RD}}{\overline{{\small \gamma }}_{I}}+\lambda
_{RP}\right) ^{L_{D}m_{RD}+m_{RP}-j-1}}\right] .  \notag
\end{eqnarray}%
\hrulefill {} \vspace*{4pt}
\end{figure*}
\begin{equation}
\mathcal{M}_{1}^{(h,l)}\left( z\right) =G_{2,3}^{2,2}\left( z\left\vert 
\begin{array}{c}
-h,1;- \\ 
\mu _{i,k}^{(h,l)},{L_{E_{k}}m_{S_{J}E_{k}}};0%
\end{array}%
\right. \right) ,  \label{M1}
\end{equation}%
\begin{equation}
\mathcal{M}_{2}^{(h,l)}\left( z\right) =G_{3,3}^{2,3}\left( z\left\vert 
\begin{array}{c}
\left( 1-m_{S_{J}P},\varphi _{J}\right) ,(1,0),(-h,0);- \\ 
({L_{E_{k}}m_{S_{J}E_{k}}},0),\left( \mu
_{i,k}^{(h,l)},0\right) ;(0,0)%
\end{array}%
\right. \right) ,  \label{M2}
\end{equation}%
\begin{equation}
\mu _{i,k}^{(h,l)}={{L_{E_{k}}m_{S_{i}E_{k}}}}%
-h+l,  \label{Mu_hl}
\end{equation}%
\begin{equation}
\xi _{v}=\lambda _{v}\left( \gamma -1\right) ;v=\{S_{i}R,RD\},  \label{Xi}
\end{equation}%
\begin{equation}
\mathcal{M}_{3}\left( z\right) =G_{2,2}^{2,1}\left( z\left\vert 
\begin{array}{c}
\left( 1-m_{S_{i}P},\varphi _{S_{i}}\right) ;\left( 1,0\right) \\ 
\left( 0,0\right) ,{\left( L_{R}m_{S_{i}R},0\right) };-%
\end{array}%
\right. \right) ,  \label{M3}
\end{equation}%
\begin{equation}
\Omega _{h}^{(i,k)}=\binom{{L_{E_{k}}m_{S_{i}E_{k}}}{\small
-1}}{{\small h}}\lambda _{S_{i}E_{k}}^{%
{L_{E_{k}}m_{S_{i}E_{k}}}-h-1},  \label{Omega_h}
\end{equation}%
\begin{equation}
\alpha _{i}^{(k,J)}=\frac{{\small \beta }_{k}^{(J)}\lambda _{S_{i}R}^{%
{L_{R}m_{S_{i}R}}}}{\Gamma \left( %
{L_{E_{k}}m_{S_{i}E_{k}}}\right) \Gamma (%
{L_{R}m_{S_{i}R}})},  \label{Alpha_ikJ}
\end{equation}%
\begin{equation}
\beta _{k}^{(J)}=\frac{1}{\Gamma \left( %
{L_{E_{k}}m_{S_{J}E_{k}}}\right) \Gamma (m_{S_{J}P})},
\label{BetakJ}
\end{equation}%
\begin{equation}
\Upsilon _{l}^{(i)}=\binom{{L_{R}m_{S_{i}R}}-1}{{\small l}}%
\gamma ^{l}\left( \gamma -1\right) ^{{L_{R}m_{S_{i}R}}-1-l},
\label{Y_l}
\end{equation}

\begin{equation}
\mathcal{B}_{j}=\binom{L_{D}m_{RD}-1}{j}\left( \gamma -1\right)
^{L_{D}m_{RD}-1-j},  \label{Bj}
\end{equation}

where $G_{p,q}^{m,n}\left( z\left\vert 
\begin{array}{c}
(a_{l})_{l\leq p} \\ 
(b_{u})_{u\leq q}%
\end{array}%
\right. \right) $ denotes the Meijer's G-function \cite[Eq. (9.301)]{table}, 
$G_{p,q}^{m,n}\left( z\left\vert 
\begin{array}{c}
(a_{l},b_{l})_{l\leq p} \\ 
(c_{u},d_{u})_{u\leq q}%
\end{array}%
\right. \right) $ accounts for the upper incomplete Meijer's G-function \cite%
[Eq. (1.1.1)]{Kilbas2004}, and $\Gamma \left( .,.\right) $ denotes the upper
incomplete Gamma function \cite[Eq. (8.350.2)]{table}. 
\begin{IEEEproof}
	The proof is provided in Appendix A.
\end{IEEEproof}

\subsection{Asymptotic Secrecy Outage Probability}

In this subsection, we provide an asymptotic analysis of the derived
closed-form expressions of the SOP. The expressions given in (\ref%
{SOP_1k_final}), (\ref{SOP1_ik_final}), and (\ref{SOP2_final}) can be
approximated for SNR regime by considering $\overline{\gamma }%
_{P}\rightarrow \infty.$

\begin{theorem}
The Asymptotic expression of the SOP in the absence of a jammer is given by (\ref{SOP_asymptotic_final_without_jammer}) as shown in the next page, while it is expressed in the presence of a jamming signal depending on various cases as follows
\begin{itemize}
\item ${L_{R}m_{S_{i}R}}<{L_{E_{k}}m_{S_{J}E_{k}}}$
\begin{eqnarray}
SOP^{(i,k,J)} &\sim &1-\prod_{k=1}^{M}\mathcal{A}_{{\small RE}_{k}{\small %
,RD,D}}\left( 1\right)  \label{Asymptotic_SOPiJ_case1} \\
&&-\frac{\sum_{k=1}^{M}\prod_{\underset{j\neq k}{j=1}}^{M}\mathcal{A}_{%
{\small RE}_{j}{\small ,RD,D}}\left( 1\right) }{\overline{{\small \gamma }}%
_{I}}  \notag \\
&&\times \mathcal{A}_{{\small RE}_{k}{\small ,RD,R,D}}\left( 1-\mathcal{P}%
\left( S_{i}R\right) \mathcal{C}_{1}^{(i,k,J)}\right),  \notag
\end{eqnarray}

\item ${L_{R}m_{S_{i}R}}>%
{L_{E_{k}}m_{S_{J}E_{k}}}$ 
\begin{eqnarray}
SOP^{(i,k,J)} &\sim &1-\prod_{k=1}^{M}\mathcal{A}_{{\small RE}_{k}{\small %
,RD,D}}\left( 1\right)  \label{Asymptotic_SOPiJ_case2} \\
&&-\frac{\sum_{k=1}^{M}\prod_{\underset{j\neq k}{j=1}}^{M}\mathcal{A}_{%
{\small RE}_{j}{\small ,RD,D}}\left( 1\right) }{\overline{{\small \gamma }}%
_{I}}  \notag \\
&&\times \mathcal{A}_{{\small RE}_{k}{\small ,RD,R,D}}\left( 1-\mathcal{P}%
\left( S_{J}E_{k}\right) \mathcal{C}_{2}^{(i,k,J)}\right) ,  \notag
\end{eqnarray}

\item ${L_{R}m_{S_{i}R}}=%
{L_{E_{k}}m_{S_{J}E_{k}}}=1$ 
\begin{align}
SOP^{(i,J)}& \sim 1-\prod_{k=1}^{M}\mathcal{A}_{{\small RE}_{k}{\small ,RD,D}%
}\left( 1\right) +\frac{\log \left( \overline{{\small \gamma }}_{I}\right) }{%
\overline{{\small \gamma }}_{I}}  \label{Asymptotic_SOPiJ_case3} \\
& \times \sum_{k=1}^{M}\prod_{\underset{j\neq k}{j=1}}^{M}\mathcal{A}_{%
{\small RE}_{j}{\small ,RD,D}}\left( 1\right) \mathcal{A}_{{\small RE}_{k}%
{\small ,RD,D}}\left( 1\right) \mathcal{C}_{3}^{(i,k,J)}.  \notag
\end{align}

\item ${L_{R}m_{S_{i}R}}=%
{L_{E_{k}}m_{S_{J}E_{k}}}$ and $%
{L_{E_{k}}m_{S_{J}E_{k}}}>1$ 
\begin{align}
SOP^{(i,J)}& \sim 1-\prod_{k=1}^{M}\mathcal{A}_{{\small RE}_{k}{\small ,RD,D}%
}\left( 1\right) -\frac{1}{\overline{{\small \gamma }}_{I}}%
\sum_{k=1}^{M}\prod_{\underset{j\neq k}{j=1}}^{M}  \notag \\
& \times \mathcal{A}_{{\small RE}_{j}{\small ,RD,D}}\left( 1\right) \mathcal{%
A}_{{\small RE}_{k}{\small ,RD,R,D}},  \label{Asymptotic_SOPiJ_case4}
\end{align}%
where $\mathcal{P}\left( q\right) =1-sgn\left( {L_{v}m_{q}}%
-1\right) ,$ $q=\{S_{i}R,S_{J}E_{k}\}$, $sgn$ stands for sign function $%
\mathcal{A}_{\bullet {\small ,\bullet ,\bullet }}\left( \bullet \right) $, $%
\mathcal{A}_{\bullet {\small ,\bullet ,\bullet ,\bullet }},$ and $\left( 
\mathcal{C}_{l}^{(i,k,J)}\right) _{l=\{1,2,3\}}$ are defined in (\ref{A1}), (%
\ref{A2}), and (\ref{C1})-(\ref{C3}), respectively.
\end{itemize}
\end{theorem}

\begin{figure*}[t]
\setcounter{mytempeqncnt}{\value{equation}} \setcounter{equation}{49}
\par
\begin{eqnarray}
SOP^{(i)} &\sim &1-\prod_{k=1}^{M}\mathcal{A}_{{\small RE}_{k}{\small ,RD,D}%
}\left( 1\right) \mathcal{A}_{{\small S}_{i}{\small E}_{k}{\small ,S}_{i}%
{\small R,R}}\left( 1\right) -\frac{1}{\overline{{\small \gamma }}_{I}}%
\sum_{k=1}^{M}\prod_{\underset{j\neq k}{j=1}}^{M}\mathcal{A}_{{\small RE}_{j}%
{\small ,RD,D}}\left( 1\right) \mathcal{A}_{{\small S}_{i}{\small E}_{j}%
{\small ,S}_{i}{\small R,R}}\left( 1\right)
\label{SOP_asymptotic_final_without_jammer} \\
&&\times \left( \mathcal{A}_{{\small RE}_{k}{\small ,RD,D}}\left( 1\right) 
\mathcal{A}_{{\small S}_{i}{\small E}_{k}{\small ,S}_{i}{\small R,S}_{i},R}+%
\mathcal{A}_{{\small S}_{i}{\small E}_{k}{\small ,S}_{i}{\small R,R}}\left(
1\right) \mathcal{A}_{{\small RE}_{k}{\small ,RD,R,D}}\right) ,  \notag
\end{eqnarray}%
\hrulefill{} \vspace*{4pt}
\end{figure*}
\begin{figure*}[t]
\setcounter{mytempeqncnt}{\value{equation}} \setcounter{equation}{50}
\par
\begin{equation}
\mathcal{A}_{e,c,v}\left( y\right) =\frac{1}{\Gamma \left( %
{L_{v}m_{e}}\right) \Gamma (m_{c})}G_{2,2}^{1,2}\left( \frac{%
\lambda _{e}}{\lambda _{c}\gamma }\left\vert 
\begin{array}{c}
-{L_{v}m_{c}}+y,1;- \\ 
{L_{v}m_{e}};0%
\end{array}%
\right. \right) ,e=\{S_{i}E_{k},RE_{k}\},c=\{S_{i}R,RD\},y=\{1,2\},%
{v=\{R,D\}}  \label{A1}
\end{equation}%
\hrulefill {} \vspace*{4pt}
\end{figure*}
\begin{figure*}[t]
\setcounter{mytempeqncnt}{\value{equation}} \setcounter{equation}{51}
\par
\begin{equation}
\mathcal{A}_{e,c,u,v}=\left\{ 
\begin{array}{c}
\frac{\lambda _{c}\left( L_{v}m_{c}-1\right) \mathcal{A}_{e,c,v}\left(
2\right) }{\Gamma (m_{uP})}\left[ \delta \gamma \left( m_{uP},\varphi
_{u}\right) +\frac{\Gamma \left( m_{uP}+1,\varphi _{u}\right) }{\lambda _{uP}%
}\right] \\ 
-\frac{\xi _{c}\mathcal{A}_{e,c,v}\left( 1\right) }{\Gamma (m_{uP})}\left[
\delta \left( \gamma \left( m_{uP},\varphi _{u}\right) +\varphi
_{R}^{m_{uP}-1}e^{-\varphi _{u}}\right) +\frac{m_{uP}}{\lambda _{uP}}\Gamma
\left( m_{uP},\varphi _{u}\right) \right]%
\end{array}%
\right\} ,\text{ }u=\{S_{i},R\}, {v=\{R,D\}}  \label{A2}
\end{equation}%
\hrulefill {} \vspace*{4pt}
\end{figure*}
\begin{figure*}[t]
\setcounter{mytempeqncnt}{\value{equation}} \setcounter{equation}{52}
\par
\begin{equation}
\mathcal{C}_{1}^{(i,k,J)}=\frac{1}{\Gamma \left( m_{S_{i}P}\right) }\left\{ 
\begin{array}{c}
\left[ \gamma \left( m_{S_{i}P},\varphi _{S_{i}}\right) \sigma _{i}^{%
{L_{R}m_{S_{i}R}}}+\frac{{\small \Gamma }\left( m_{S_{i}P}+%
{L_{R}m_{S_{i}R}},\varphi _{S_{i}}\right) }{\lambda
_{S_{i}P}^{{L_{R}m_{S_{i}R}}}}\right] \frac{\xi _{S_{i}R}^{%
{L_{R}m_{S_{i}R}}}}{{L_{R}m_{S_{i}R}}\Gamma (%
{L_{R}m_{S_{i}R})}}+\gamma \alpha _{i}^{(k,J)}\sum_{l=0}^{%
{L_{R}m_{S_{i}R}}-1}\Upsilon _{l} \\ 
\times \left[ \frac{\gamma \left( m_{S_{i}P},\varphi _{S_{i}}\right) }{%
\sigma _{i}^{-{L_{R}m_{S_{i}R}}+l+1}}+\frac{\Gamma \left(
m_{S_{i}P}+{L_{R}m_{S_{i}R}}-l-1,\varphi _{S_{i}}\right) }{%
\lambda _{S_{i}P}^{{L_{R}m_{S_{i}R}}-l-1}}\right] \frac{%
{\small \Gamma }\left( {L_{E_{k}}m_{S_{J}E_{k}}}-l{\small -1}%
\right) {\small \Gamma }\left( {L_{E_{k}}m_{S_{i}E_{k}}}%
+l+1\right) }{l+1} \\ 
\times \left[ \gamma \left( m_{S_{J}P},\varphi _{J}\right) \left( \frac{%
\delta \lambda _{S_{J}E_{k}}}{\lambda _{S_{i}E_{k}}}\right) ^{l+1}+{\small %
\Gamma }\left( m_{S_{J}P}+l+1,\varphi _{J}\right) \left( \varsigma
_{i}^{(k,J)}\right) ^{l+1}\right]%
\end{array}%
\right\} ,  \label{C1}
\end{equation}%
\hrulefill{} \vspace*{4pt}
\end{figure*}
\begin{figure*}[t]
\setcounter{mytempeqncnt}{\value{equation}} \setcounter{equation}{53}
\par
\begin{eqnarray}
\mathcal{C}_{2}^{(i,k,J)} &\sim &\frac{\gamma ^{%
{L_{R}m_{S_{i}R}}}\alpha _{i}^{(k,J)}}{%
{L_{E_{k}}m_{S_{J}E_{k}}}}\left[ \gamma \left(
m_{S_{J}P},\varphi _{J}\right) \left( \frac{\delta \lambda _{S_{J}E_{k}}}{%
\lambda _{S_{i}E_{k}}}\right) ^{{L_{E_{k}}m_{S_{J}E_{k}}}}+%
{\small \Gamma }\left( m_{S_{J}P}+{L_{E_{k}}m_{S_{J}E_{k}}}%
,\varphi _{J}\right) \left( \varsigma _{i}^{(k,J)}\right) ^{%
{L_{E_{k}}m_{S_{J}E_{k}}}}\right]  \label{C2} \\
&&\times \sum_{h{\small =0}}^{{L_{E_{k}}m_{S_{i}E_{k}}}{%
\small -1}}\frac{\Omega _{h}^{(i,k)}\Gamma \left( %
{L_{E_{k}}m_{S_{i}E_{k}}}+{L_{R}m_{S_{i}R}}-%
{L_{E_{k}}m_{S_{J}E_{k}}}-h-1\right) {\small \Gamma }\left(%
{L_{E_{k}}m_{S_{J}E_{k}}}+h+1\right) }{\left( \varpi
_{i}^{(k)}\right) ^{{L_{E_{k}}m_{S_{i}E_{k}}}+%
{L_{R}m_{S_{i}R}}-{L_{E_{k}}m_{S_{J}E_{k}}}%
-h-1}},  \notag
\end{eqnarray}%
\hrulefill{} \vspace*{4pt}
\end{figure*}
\begin{figure*}[t]
\setcounter{mytempeqncnt}{\value{equation}} \setcounter{equation}{54}
\par
\begin{eqnarray}
\mathcal{C}_{3}^{(i,k,J)} &=&\frac{\gamma ^{{L_{R}m_{S_{i}R}}%
}\alpha _{i}^{(k,J)}}{\varpi ^{{L_{R}m_{S_{i}R}}-%
{L_{E_{k}}m_{S_{J}E_{k}}}}}\left[ \gamma \left(
m_{S_{J}P},\varphi _{J}\right) \left( \frac{\delta \lambda _{S_{J}E_{k}}}{%
\lambda _{S_{i}E_{k}}}\right) ^{{L_{E_{k}}m_{S_{J}E_{k}}}%
}+\left( \varsigma _{i}^{(k,J)}\right) ^{%
{L_{E_{k}}m_{S_{J}E_{k}}}}{\small \Gamma }\left( m_{S_{J}P}+%
{L_{E_{k}}m_{S_{J}E_{k}}},\varphi _{J}\right) \right]
\label{C3} \\
&&\frac{\left( -1\right) ^{{L_{E_{k}}m_{S_{J}E_{k}}}-%
{L_{R}m_{S_{i}R}}}{\small \Gamma }\left( %
{L_{E_{k}}m_{S_{i}E_{k}}}+%
{L_{E_{k}}m_{S_{J}E_{k}}}\right) }{%
{L_{E_{k}}m_{S_{J}E_{k}}}}.  \notag
\end{eqnarray}%
\hrulefill{} \vspace*{4pt}
\end{figure*}
\begin{IEEEproof}
	The proof is provided in Appendix B.
\end{IEEEproof}

\section{Results and discussion}

In this section, we validate the derived analytical results through Monte
Carlo simulation by generating $10^{6}$ Gamma-distributed random variables. 
{The setting parameters of the simulation are summarized in Table 1. Indeed, the values of fading severity parameter $ m_{\bullet}$ have
been chosen such that the wiretap channel is better than the legitimate one.
Moreover, their values are taken integer in the range {$2$..$5$} similarly
to \cite{lei2017secrecy} and \cite{Zhong2011}. On the other hand, the
average SNR, which is inversely proportional to $\lambda_{\bullet}$, the
legitimate link is considered better than the one of the wiretap channel}.
It is worthwhile that these parameters are associated with all figures
except those indicating other values. As one can see in Figs. 2-5, all
closed-form and simulation curves are perfectly matched for considered
parameters' values. 
\begin{table}[tbp]
\caption{{Simulation parameters.}}%
\begin{centering}
		\begin{tabular}{c|c|c|c|c|c}
			\hline
			\hline
			Parameter &$ M $&$ N $& $ \lambda_{S_{i}R} $ & $ \lambda_{S_{i}P} $ & $ \lambda_{S_{i}E_{k}} $  \tabularnewline
			\hline
			value  & 3 & 4 & 0.1 & 0.3 & 0.6   \tabularnewline
			\hline
			Parameter & $ \lambda_{RE_{k}} $ & $ \lambda_{RP} $ & $ \lambda_{RD} $ & $ m_{S_{i}R}$ & $ m_{S_{i}P} $   \tabularnewline
			\hline
			value & 0.6 & 0.2 & 0.1 & 2 &  3 \tabularnewline
			\hline
			Parameter & $ m_{S_{i}E_{k}} $ & $m_{RD}$ & $m_{RE_{k}}$ & $m_{RP}$&    \tabularnewline
			\hline
			value & 5 & 2 & 4 & 3 &    \tabularnewline
			\hline
			\hline
		\end{tabular}
		\par\end{centering}
\centering{}
\end{table}

Fig. 2 and Fig. 3 depict closed-form and asymptotic expressions for the SOP
versus $\overline{\gamma }_{I}$ for various values of antennas' numbers in
both the presence and absence of a friendly jammer cases, respectively. As
stated in remark $1$, It can be noticed that the greater $\overline{\gamma }%
_{I}$, the smaller the SOP. Interestingly, above a certain threshold of $%
\overline{\gamma }_{I}$ the SOP becomes steady this can be obviously
justified from (\ref{P_Si}) and (\ref{P_R}) that above that threshold, both
sources and relay will always transmit with their maximum powers.
Consequently, the legitimate and wiretap capacities of each hop remain
constant, leading to a constant value of SOP. Interestingly, by comparing
the SOP values in the two aforementioned figures, one can ascertain that
better secrecy is achieved by using a friendly jammer. In addition, the
asymptotic curves are plotted under the considered fading severity values
(i.e., $m_{S_{i}R}=2$, $m_{S_{J}E_{k}}=5$) from Eqs. (\ref%
{Asymptotic_SOPiJ_case1}), (\ref{A1})-(\ref{C1}). Clearly, the asymptotic
curves match with the closed-form ones in high SNR regime. 
\begin{figure}[tbp]
\begin{centering}
		\includegraphics[width=98mm,height=60mm]{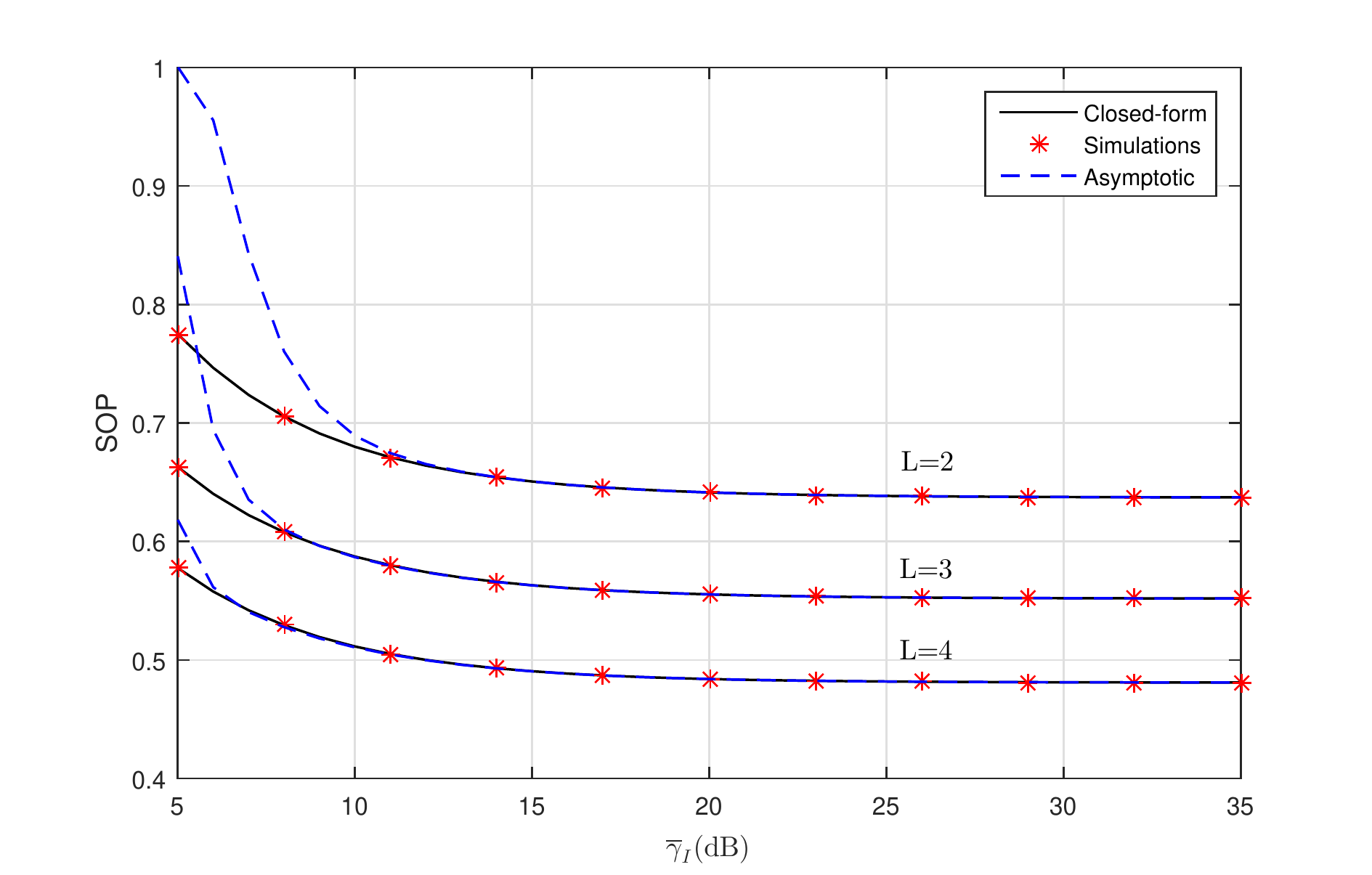}
		\par\end{centering}
\centering{}
\caption{SOP vs $\overline{\protect\gamma }_{I}$ for different values of
antennas at the destination in the presence of a friendly jammer for $%
\protect\eta =\protect\sigma _{i}=\protect\delta =0.1$ {and $L_{R}=L_{E_{k}}=L_{D}=L$}.}
\label{Fig2}
\end{figure}
\begin{figure}[tbp]
\begin{centering}
		\includegraphics[width=98mm,height=60mm]{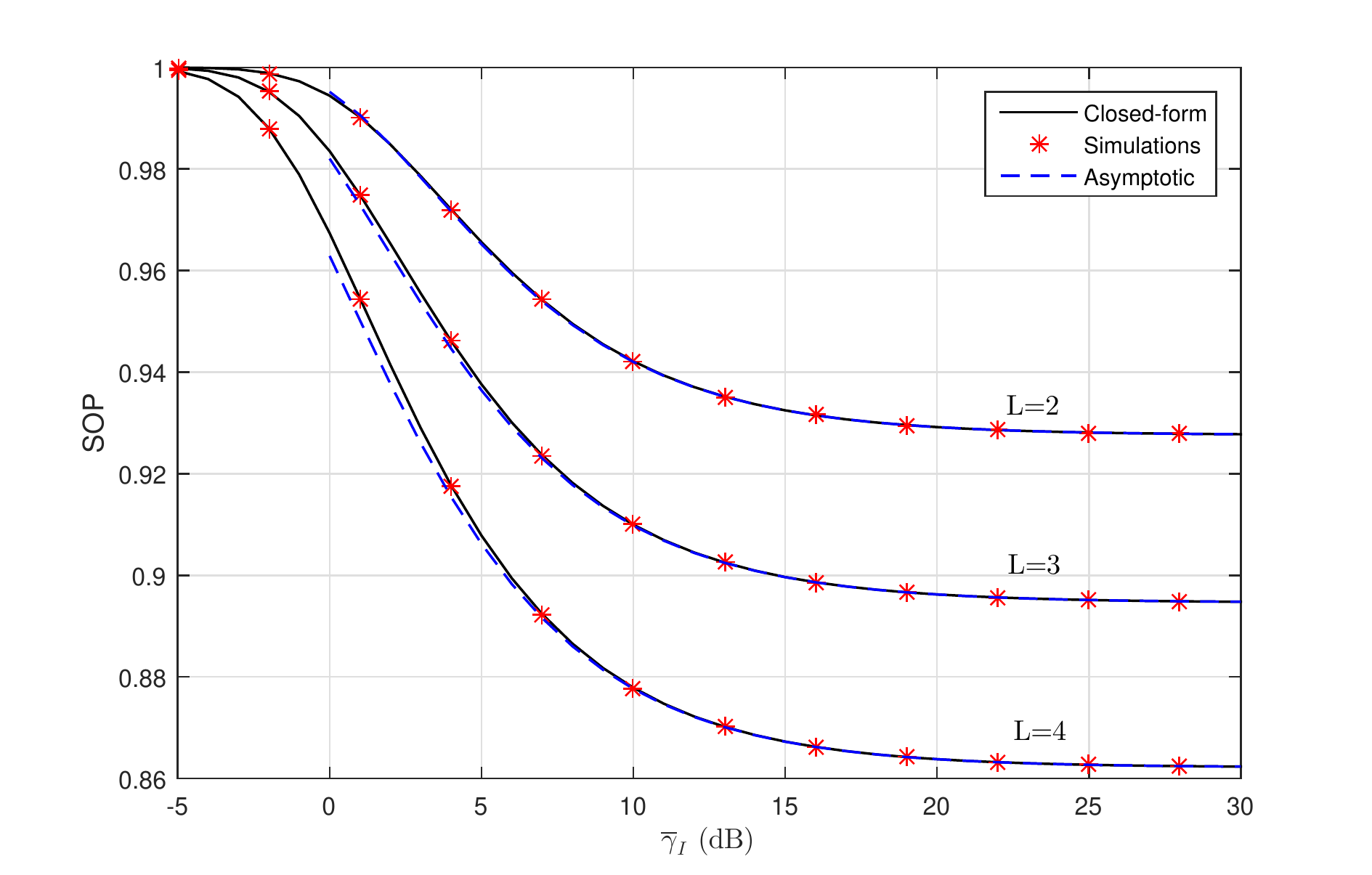}
		\par\end{centering}
\centering{}
\caption{SOP vs $\overline{\protect\gamma }_{I}$ in the absence of a
friendly jammer for $\protect\sigma_{i}=\protect\delta =0.1$ {and $L_{R}=L_{E_{k}}=L_{D}=L$}.}
\end{figure}

Fig. 4 illustrates the SOP versus $\overline{\gamma }_{S_{J}}$ for numerous
values of branches' number $L_{D}$ at the receiver $D$. Again, as indicated
in remark $1$, one can realize that the higher $\overline{\gamma }_{S_{J}}$
and $L_{D},$ the smaller the SOP and therefore the system's security gets
improved. 
\begin{figure}[tbp]
\begin{centering}
		\includegraphics[width=98mm,height=60mm]{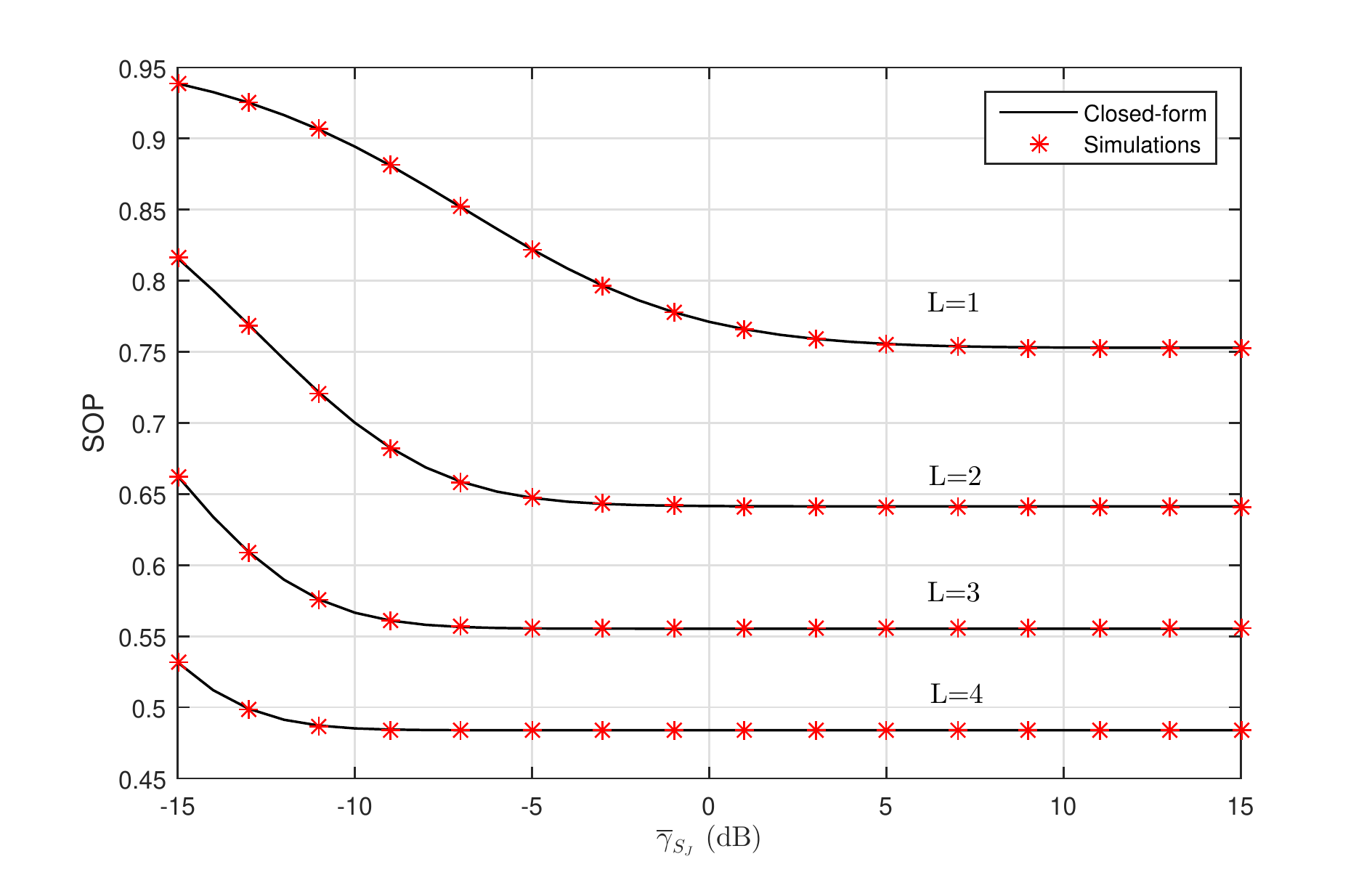}
		\par\end{centering}
\centering{}
\caption{SOP vs $\overline{\protect\gamma }_{S_{J}}$ for different values of
antennas at the destination {for} $\overline{\protect\gamma }_{I}=\overline{%
\protect\gamma }_{S_{i}}=\overline{\protect\gamma }_{R}=20$ dB {and $L_{R}=L_{E_{k}}=L_{D}=L$}.}
\end{figure}

Fig. 5 and Fig. 6 show the SOP as a function of the number of eavesdroppers $%
M$ for different values of $\overline{\gamma }_{S_{J}}$ and by considering
both cases i.e., presence and absence of jammer. One can observe that the
smaller $\overline{\gamma }_{S_{J}}$ or the greater $M$ the worst is the
system's secrecy as highlighted in remark $1$ and $2,$ respectively. \ In
addition, introducing a jamming signal improves significantly the secrecy
performance for high values of $\overline{\gamma }_{S_{J}}$ or in the
presence of small numbers of eavesdroppers. 
\begin{figure}[tbp]
\begin{centering}
		\includegraphics[width=98mm,height=60mm]{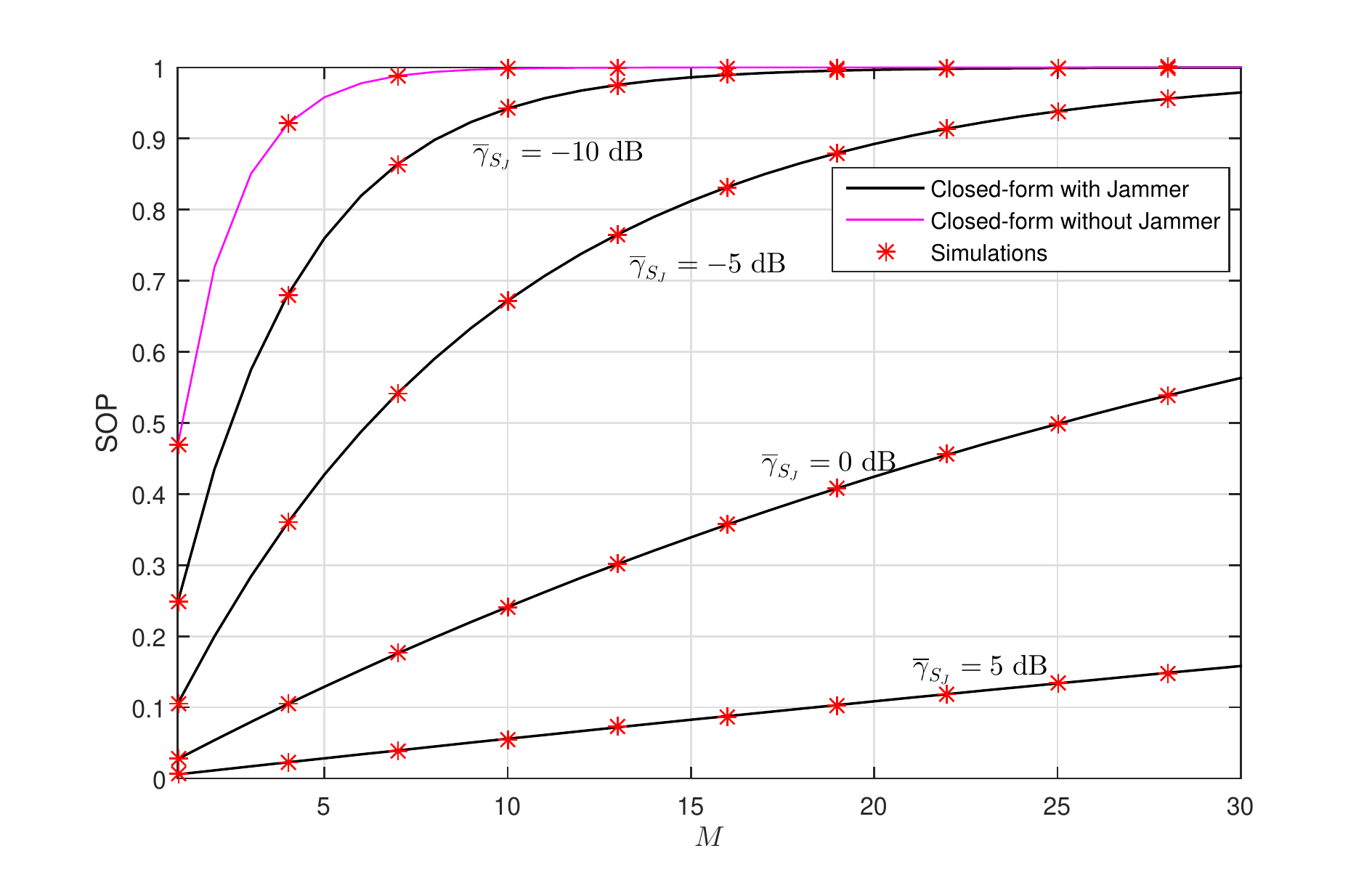}
		\par\end{centering}
\centering{}
\caption{SOP vs number of eavesdroppers and different values $\overline{%
\protect\gamma }_{S_{J}}$ for $L_{D}=4$ {and} $\overline{\protect\gamma }_{I}=%
\overline{\protect\gamma }_{S_{i}}=\overline{\protect\gamma }_{R}=20$ dB.}
\end{figure}
\begin{figure}[tbp]
\begin{centering}
		\includegraphics[width=98mm,height=60mm]{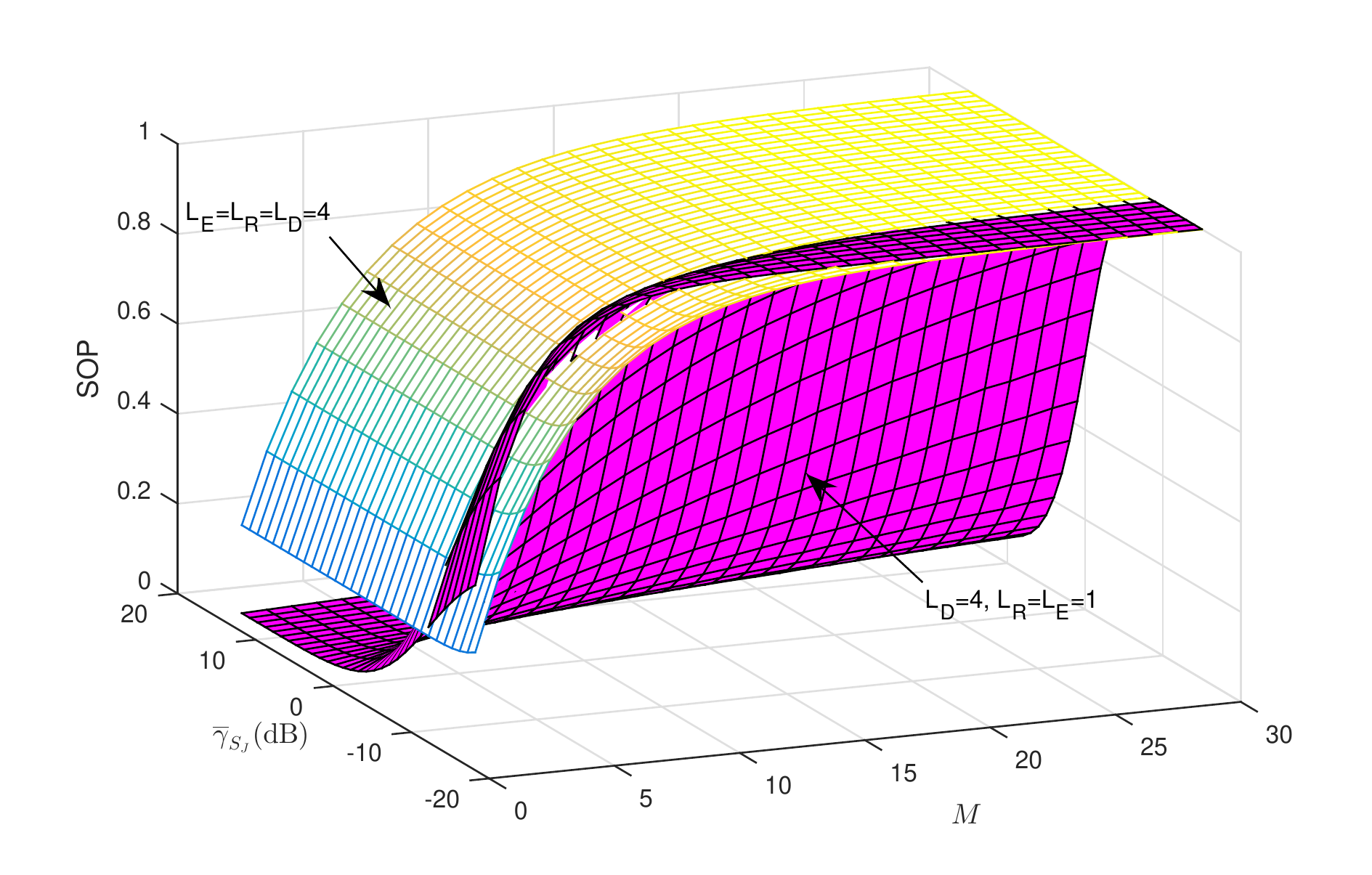}
		\par\end{centering}
\centering{}
\caption{SOP vs number of eavesdroppers and $\overline{\protect\gamma }%
_{S_{J}}$ for $\overline{\protect\gamma }_{I}=\overline{\protect%
\gamma }_{S_{i}}=\overline{\protect\gamma }_{R}=20$ dB.}
\end{figure}
{Fig. 7 depicts the SOP as a function of the number of
eavesdroppers by considering the presence and absence of a friendly jammer.
It is worth mentioning that better security is obviously achieved for the
case of presence of jammer and multi-antenna nodes, while the scenario of
the absence of jammer and legitimate nodes equipped with a single antenna is the
worst case. For this reason, our aim here is to investigate if the security
gets enhanced when having artificial noise and legitimate nodes with a single
antenna or the scenario of the absence of jammer and all legitimate nodes are
equipped with multiple antennas. One can obviously notice that the system's
security is improved when diversity is used at the legitimate nodes.
Additionally, in the presence of an important number of eavesdroppers, the
friendly jammer does not contribute to the enhancement of the system's
security.} 
\begin{figure}[tbp]
\begin{centering}
		\includegraphics[width=98mm,height=60mm]{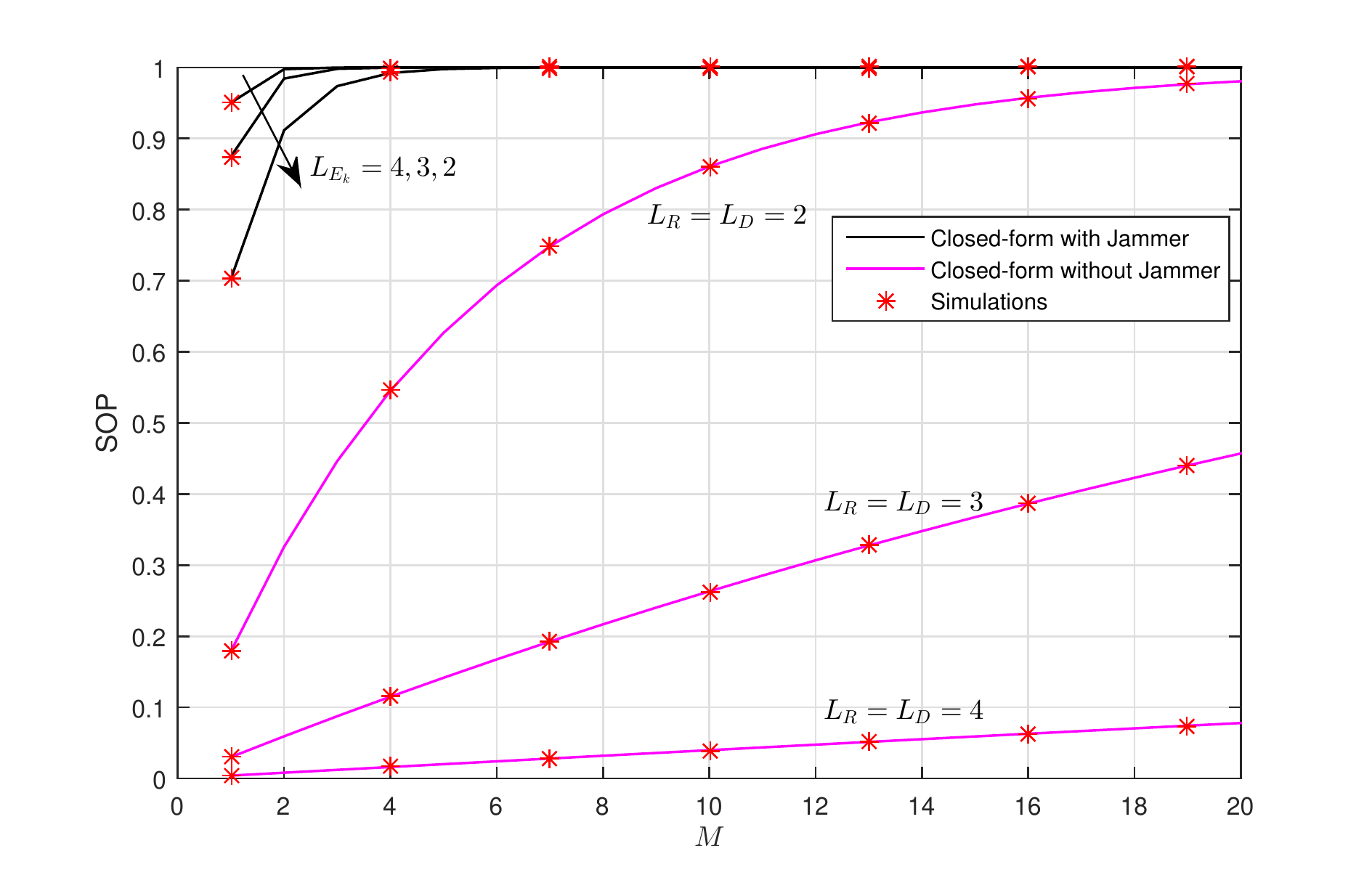}
		\par\end{centering}
\centering{}
\caption{SOP vs the number of eavesdroppers in the presence and absence of
friendly jammer and different numbers of antennas {for }$\overline{\protect\gamma }%
_{S_{J}}=20$ dB.}
\end{figure}

\section{Conclusions}

In this paper, the physical layer security of a dual-hop underlay uplink CRN
operating under Nakagami-$m$ fading channels {was} investigated. We considered
multiple sources communicating, in turn, with the base station through a
relay in the presence of several eavesdroppers attempting to overhear the
communication channels. 
{All receivers, i.e., legitimates and
wiretappers, {were} assumed to be equipped with multiple antennas and perform
the MRC technique.} Closed-form and asymptotic expressions for the SOP under
various cases of fading parameters' values were derived by considering two
scenarios namely, (i) presence and (ii) absence of a friendly jammer. The
obtained results showed that the best secrecy is achieved in the presence of
a small number of eavesdroppers when increasing the transmit power of the
SUs', the number of antennas at the legitimate receiver and the maximum
tolerated interference power at the PU as well. 
{Interestingly, we showed that equipping the legitimate nodes
by multiple antennas leads to a noticeable enhancement of the system's
security rather than sending an artificial noise.} As future work, we intend
to investigate the impact of NOMA jointly with the key parameters considered
in this work on the system's secrecy. 
{We also intend to
consider the case of amplify-and-forward relaying {protocol}
and investigate the impact of numerous jammer selection policies on the
overall system's security.}

\section*{Appendix A: proof of theorem 1}

\subsection{Expression of $SOP$ at the First Hop}

The SOP at the first hop in the absence and presence of a friendly jammer is
given, respectively, by 
\begin{equation}
SOP_{1}^{(i,k)}=1-\gamma \int_{x=0}^{\infty }f_{g_{S_{i}P}}(x)\Xi
_{2}^{(i,k)}\left( x\right) dx,  \label{SOP1_without_jammer}
\end{equation}
\begin{equation}
SOP_{1}^{(i,k,J)}=1-\gamma \int_{x=0}^{\infty }f_{g_{S_{i}P}}(x)\Xi
_{1}^{(i,k,J)}\left( x\right) dx,  \label{SOP1_exp2}
\end{equation}
where (\ref{SOP1_exp2}) and (\ref{SOP1_without_jammer}) hold by using
integration by parts on \cite[Eq. (33)]{bouabdellah2018secrecy} , with 
\begin{equation}
\Xi _{1}^{(i,k,J)}\left( x\right) =\int_{0}^{\infty }f_{\gamma
_{R}^{(i)}|g_{S_{i}P}=x}\left( \gamma y+\gamma -1\right) F_{\gamma
_{1E}^{(i,k,J)}|g_{S_{i}P}=x}(y)dy,  \label{E1_exp1}
\end{equation}
and%
\begin{equation}
\Xi _{2}^{(i,k)}\left( x\right) =\int_{0}^{\infty }f_{\gamma
_{R}^{(i)}|g_{S_{i}P}=x}\left( \gamma y+\gamma -1\right) F_{\gamma
_{1E}^{(i,k)}|g_{S_{i}P}=x}(y)dy,  \label{E1_exp1_ik}
\end{equation}%
and $\gamma $ is being defined in {Theorem 1}.

\begin{itemize}
\item Conditional CDF of $\gamma _{R}^{(i)}$
\end{itemize}

The CDF of $\gamma _{R}^{(i)}$ for a given $g_{S_{i}P}$ can be expressed as 
\begin{align}
F_{\gamma _{R}^{(i)}|g_{S_{i}P}=x}\left( z\right) & =\Pr \left( \min \left( 
\overline{{\small \gamma }}_{S_{i}},\frac{\overline{{\small \gamma }}_{I}}{x}%
\right) {\mathcal{E}_{S_{i}R}}\leq z\right)  \label{F_gammaR}
\\
& ={F_{\mathcal{E}_{S_{i}R}}}\left( \frac{z}{\Phi \left(
x\right) }\right) ,  \notag
\end{align}%
where 
{$\mathcal{E}_{S_{i}R}=$
$\sum_{u=1}^{L_{R}}g_{S_{i}R_{u}},$} $\Phi \left( x\right) =$ $\overline{%
{\small \gamma }}_{S_{i}}$ for $x\leq \overline{{\small \gamma }}_{I}/%
\overline{{\small \gamma }}_{S_{i}}$ and $\Phi \left( x\right) =$ $\overline{%
{\small \gamma }}_{I}/x$ for $x>\overline{{\small \gamma }}_{I}/\overline{%
{\small \gamma }}_{S_{i}}.$

\subsubsection{SOP at the {F}irst {H}op with the {A}bsence of a {J}amming {S}ignal}

The conditional CDF of $\gamma _{1E}^{(i,k)}$ can be expressed as 
\begin{align}
F_{\gamma _{1E}^{(i,k)}|g_{S_{i}P}=x}(y)& =\Pr \left( \min \left( \overline{%
{\small \gamma }}_{S_{i}},\frac{\overline{{\small \gamma }}_{I}}{x}\right) %
{\mathcal{E}_{S_{i}E_{k}}}\leq y\right)  \notag \\
& ={F_{\mathcal{E}_{S_{i}E_{k}}}}\left( \frac{y}{\Phi \left(
x\right) }\right).  \label{Fgamma1E_noJammaer}
\end{align}
where {$\mathcal{E}_{S_{i}E_{k}}=%
\sum_{u=1}^{L_{R}}g_{S_{i}E_{k}^{(u)}}.$}

{It is worth mentioning that for i.i.d Nakagami-$m$ channels,
$\mathcal{E}_{S_{i}R}$ and $\mathcal{E}_{S_{i}E_{k}}$ are Gamma distributed
with shape and scale parameters ${L_{R}m_{S_{i}R}}$ and
$\lambda _{S_{i}R}$, ${L_{E_{k}}m_{S_{i}E_{k}}}$ and $\lambda
_{S_{i}E_{k}},$ respectively.}

Substituting (\ref{F_gammaR}) and (\ref{Fgamma1E_noJammaer}) into (\ref%
{E1_exp1_ik}), and using \cite[Eqs. (06.06.26.0004.01), (07.34.21.0088.01]%
{Wolfram} yields

\begin{align}
\Xi _{2}^{(i,k)}\left( x\right) & =\frac{\lambda _{S_{i}R}^{%
{L_{R}m_{S_{i}R}}}e^{-\frac{\xi _{S_{i}R}}{\Phi \left(
x\right) }}}{\Phi ^{{L_{R}m_{S_{i}R}}}\left( x\right) \Gamma
\left( {L_{E_{k}}m_{S_{i}E_{k}}}\right) \Gamma (%
{L_{R}m_{S_{i}R}})}\sum_{l=0}^{%
{L_{R}m_{S_{i}R}}-1}\Upsilon _{l}^{(i)}
\label{E1_final_noJammer} \\
& \times \left( \frac{\lambda _{S_{i}R}\gamma }{\Phi \left( x\right) }%
\right) ^{-l-1}G_{2,2}^{1,2}\left( \frac{\lambda _{S_{i}E_{k}}}{\lambda
_{S_{i}R}\gamma }\left\vert 
\begin{array}{c}
-l,1;- \\ 
{L_{E_{k}}m_{S_{i}E_{k}}};0%
\end{array}%
\right. \right) ,  \notag
\end{align}%
where $\Upsilon _{l}^{(i)}$ is defined in (\ref{Y_l}).

Now, replacing (\ref{E1_final_noJammer}) into (\ref{SOP1_without_jammer}),
one can obtain

\begin{align}
SOP_{1}^{(i,k)}& =1-\frac{\lambda _{S_{i}R}^{{L_{R}m_{S_{i}R}}%
}\gamma }{\Gamma \left( {L_{E_{k}}m_{S_{i}E_{k}}}\right)
\Gamma ({L_{R}m_{S_{i}R}})}\sum_{l=0}^{%
{L_{R}m_{S_{i}R}}-1}\frac{\Upsilon _{l}^{(i)}}{\left( \gamma
\lambda _{S_{i}R}\right) ^{l+1}}  \notag \\
& \times \mathcal{H}_{1}G_{2,2}^{1,2}\left( \frac{\lambda _{S_{i}E_{k}}}{%
\lambda _{S_{i}R}\gamma }\left\vert 
\begin{array}{c}
-l,1;- \\ 
{L_{E_{k}}m_{S_{i}E_{k}}};0%
\end{array}%
\right. \right) ,  \label{SOP1_ik}
\end{align}%
where 
\begin{align}
& \mathcal{H}_{1}=\int_{0}^{\infty }\frac{f_{g_{S_{i}P}}(x)e^{-\frac{\xi
_{S_{i}R}}{\Phi \left( x\right) }}}{\Phi ^{{L_{R}m_{S_{i}R}}%
-l-1}\left( x\right) }dx  \label{Hi_final} \\
& \quad \quad \overset{(a)}{=}\frac{\overline{{\small \gamma }}_{I}^{-%
{L_{R}m_{S_{i}R}}+l+1}}{\Gamma \left( m_{S_{i}P}\right) }%
\left[ 
\begin{array}{c}
\frac{e^{-\frac{\sigma _{i}\xi _{S_{i}R}}{\overline{{\small \gamma }}_{I}}}}{%
\sigma _{i}^{-{L_{R}m_{S_{i}R}}+l+1}}\gamma \left(
m_{S_{i}P},\varphi _{S_{i}}\right) \\ 
+\frac{\lambda _{S_{i}P}^{m_{S_{i}P}}\Gamma \left( %
{L_{R}m_{S_{i}R}}+m_{S_{i}P}-l-1,\varphi _{S_{i}}+\frac{\xi
_{S_{i}R}}{\overline{{\small \gamma }}_{S_{i}}}\right) }{\left( \lambda
_{S_{i}P}+\frac{\xi _{S_{i}R}}{\overline{{\small \gamma }}_{I}}\right) ^{%
{L_{R}m_{S_{i}R}}+m_{S_{i}P}-l-1}}%
\end{array}%
\right] ,  \notag
\end{align}%
where step $(a)$ is obtained by replacing $\Phi \left( x\right) $ by its
values, and performing some algebraic manipulations.

Now, incorporating (\ref{Hi_final}) into (\ref{SOP1_ik}), (\ref%
{SOP1_ik_final}) is attained.

\subsubsection{SOP at the {F}irst {H}op {in} the {P}resence of a {J}amming {S}ignal}

In the presence of a friendly jammer, the CDF of $\gamma _{1E}^{(i,k,J)}$
for a given $g_{S_{i}P}$ is given by 
\begin{subequations}
\begin{align}
& F_{\gamma _{1E}^{(i,k,J)}|g_{S_{i}P}=x}(y)\overset{(a)}{=}\int_{0}^{\infty
}{F_{\mathcal{E}_{S_{i}E_{k}}}}\left( \frac{y\left(
t+1\right) }{\Phi \left( x\right) }\right)  \label{Fgamma1Ea} \\
& \quad \quad \quad \quad \quad \quad \quad \quad \times
f_{W_{k}^{(J)}}\left( t\right) dt  \notag \\
& \quad \quad \quad \quad \quad \quad \quad \overset{(b)}{=}1-\Psi
_{i}^{(k)}\left( y\right)  \label{Fgamma1Eb} \\
& \quad \quad \quad \quad \quad \quad \quad \quad \times \sum_{h=0}^{%
{L_{E_{k}}m_{S_{i}E_{k}}}-1}\binom{%
{L_{E_{k}}m_{S_{i}E_{k}}}-1}{h}\mathcal{V}^{(h)}(y),  \notag
\end{align}%
where $W_{k}^{(J)}=\min \left( \overline{\gamma }_{S_{J}},\frac{\overline{%
\gamma }_{P}}{g_{S_{J}P}}\right) {\mathcal{E}_{S_{J}E_{k}}},$ 
$%
{\mathcal{E}
_{S_{J}E_{k}}=\sum_{u=1}^{L_{E_{k}}}g_{S_{J}E_{k}^{(u)}}},$ $\mathcal{V}%
^{(h)}(y)=$ $\int_{0}^{\infty }t^{h}e^{-\frac{y\lambda _{S_{i}E_{k}}}{\Phi
\left( x\right) }t}F_{W_{k}^{(J)}}\left( t\right) dt,$ $\Psi
_{i}^{(k)}\left( y\right) =\frac{yf_{S_{i}E_{k}}\left( \frac{y}{\Phi \left(
x\right) }\right) }{\Phi \left( x\right) }.$ Here step (\ref{Fgamma1Ea})
holds using the definition (\ref{Gamma_E1}), while step (\ref{Fgamma1Eb}) is
obtained by using integration by parts alongside the Binomial formula for a
positive integer ${L_{E_{k}}m_{S_{i}E_{k}}}$. Importantly,
the derivation of the CDF of $\gamma _{1E}^{(i,k,J)}$ requires the one of $%
W_{k}^{(J)},$ given as 
\end{subequations}
\begin{align}
F_{W_{k}^{(J)}}\left( t\right) & =\Pr \left( {%
\mathcal{E}_{S_{J}E_{k}}}\leq \frac{t}{\overline{{\small \gamma }}_{S_{J}}},%
\frac{\overline{{\small \gamma }}_{I}}{g_{S_{J}P}}\geq \overline{{\small %
\gamma }}_{S_{J}}\right)  \label{F_Wik_exp1} \\
& +\Pr \left( \frac{{\mathcal{E}_{S_{J}E_{k}}}}{g_{S_{J}P}}%
\leq \frac{t}{\overline{{\small \gamma }}_{I}},\frac{\overline{{\small %
\gamma }}_{I}}{g_{S_{J}P}}\leq \overline{{\small \gamma }}_{S_{J}}\right) 
\notag \\
& ={F_{\mathcal{E}_{S_{J}E_{k}}}}\left( \frac{t}{\overline{%
{\small \gamma }}_{S_{J}}}\right) F_{g_{S_{J}P}}\left( \frac{\overline{%
{\small \gamma }}_{I}}{\overline{{\small \gamma }}_{S_{J}}}\right) +\mathcal{%
I}_{1}^{(k,J)}.  \notag
\end{align}%
where 
\begin{align}
& \mathcal{I}_{1}^{(k,J)}=\text{ }\int_{\frac{\overline{{\small \gamma }}_{I}%
}{\overline{{\small \gamma }}_{S_{J}}}}^{\infty }f_{g_{S_{J}P}}\left( \nu
\right) {F_{\mathcal{E}_{S_{J}E_{k}}}}\left( \frac{t}{%
\overline{{\small \gamma }}_{I}}\nu \right) d\nu  \label{I1} \\
& \quad \quad \overset{(a)}{=}\text{ }\int_{\mathcal{L}_{1}}\frac{{\small %
\Gamma }\left( m_{S_{J}P}-s,\varphi _{J}\right) {\small \Gamma }\left( %
{L_{E_{k}}m_{S_{J}E_{k}}}+s\right) {\small \Gamma }\left(
-s\right) }{2\pi j{\small \Gamma }\left( 1-s\right) \left( \kappa t\right)
^{s}\left( \beta _{k}^{(J)}\right) ^{-1}}{\small ds}  \notag \\
& \quad \quad =\beta _{k}^{(J)}\Delta _{k}^{(J)}\left( t\right) ,  \notag
\end{align}%
where $j=\sqrt{-1},$ $\mathcal{L}_{1}$ is a vertical line of integration
chosen such as to separate the left poles of the above integrand function
from the right ones, $\Delta _{k}^{(J)}\left( t\right) =G_{2,2}^{1,2}\left(
\kappa t\left\vert 
\begin{array}{c}
\left( 1-m_{S_{J}P},\varphi _{J}\right) ,(1,0);- \\ 
({L_{E_{k}}m_{S_{J}E_{k}}},0);(0,0)%
\end{array}%
\right. \right) $, $\kappa =\lambda _{S_{J}E_{k}}/\lambda _{S_{J}P}\overline{%
{\small \gamma }}_{I},$ $\varphi _{J}$ , and $\beta _{k}^{(J)}$ are defined
in {Theorem 1} and (\ref{BetakJ}), respectively. Step $(a)$
holds using \cite[Eq. (06.06.26.0004.01)]{Wolfram} alongside with (\ref{f_gq}%
) and (\ref{F_gq}). 
{As mentioned above, $\mathcal{E}_{S_{J}E_{k}}$ is also Gamma
distributed with parameters ${L_{E_{k}}m_{S_{J}E_{k}}}$ and $\lambda
_{S_{J}E_{k}}.$}

Substituting (\ref{I1}) into (\ref{F_Wik_exp1}), we get 
\begin{equation}
F_{W_{k}^{(J)}}\left( t\right) =\beta _{k}^{(J)}\left[ 
\begin{array}{c}
\gamma \left( {L_{E_{k}}m_{S_{J}E_{k}}},\frac{\lambda
_{S_{J}E_{k}}}{\overline{{\small \gamma }}_{S_{J}}}t\right) \\ 
\times \gamma \left( m_{S_{J}P},\varphi _{J}\right) +\Delta _{k}^{(J)}\left(
t\right)%
\end{array}%
\right] .  \label{Fwik_exp2}
\end{equation}

Now, it remains to compute $\mathcal{V}^{(h)}(y)$ so as to evaluate (\ref%
{Fgamma1Eb}). Using (\ref{Fwik_exp2}), yields 
\begin{equation}
\mathcal{V}^{(h)}(y)=\beta _{k}^{(J)}\left( \gamma \left( m_{S_{J}P},\varphi
_{J}\right) \mathcal{T}_{1}^{(h)}+\mathcal{T}_{2}^{(h)}\right) ,
\label{Fgamma1E_exp2}
\end{equation}%
where 
\begin{align}
& \mathcal{T}_{1}^{(h)}=\int_{0}^{\infty }t^{h}e^{-\frac{y\lambda
_{S_{i}E_{k}}}{\Phi \left( x\right) }t}\gamma \left( %
{L_{E_{k}}m_{S_{J}E_{k}}},\frac{\lambda _{S_{J}E_{k}}}{%
\overline{{\small \gamma }}_{S_{J}}}t\right) dt  \notag \\
& \quad \quad \overset{(a)}{=}\int_{0}^{\infty }t^{h}e^{-\frac{y\lambda
_{S_{i}E_{k}}}{\Phi \left( x\right) }t}  \notag \\
& \quad \quad \quad \times G_{1,2}^{1,1}\left( \frac{\lambda _{S_{J}E_{k}}}{%
\overline{{\small \gamma }}_{S_{J}}}t\left\vert 
\begin{array}{c}
1;- \\ 
{L_{E_{k}}m_{S_{J}E_{k}}};0%
\end{array}%
\right. \right) dt  \notag \\
& \quad \quad \overset{(b)}{=}\left( \frac{\Phi \left( x\right) }{\lambda
_{S_{i}E_{k}}y}\right) ^{h+1}\Theta _{1}^{(h)}\left( y\right) ,
\label{T1_exp_final}
\end{align}%
with $\Theta _{1}^{(h)}\left( y\right) =G_{2,2}^{1,2}\left( \frac{\theta
_{i}^{(k,J)}\Phi \left( x\right) }{y}\left\vert 
\begin{array}{c}
-h,1;- \\ 
{L_{E_{k}}m_{S_{J}E_{k}}};0%
\end{array}%
\right. \right) $ and $\theta _{i}^{(k,J)}$ is being defined in %
{Theorem 1}. The equalities $(a)$ and $(b)$ follow by using
Eqs. (06.06.26.0004.01) and (07.34.21.0088.01) of \cite{Wolfram},
respectively.

On the other hand, the term $\mathcal{T}_{2}^{(h)}$ can be expressed as

\begin{align}
\mathcal{T}_{2}^{(h)}& =\int_{0}^{\infty }t^{h}e^{-\frac{y\lambda
_{S_{i}E_{k}}}{\Phi \left( x\right) }t}\Delta _{k}^{(J)}\left( t\right) dt
\label{T2} \notag\\
& =\frac{1}{2\pi j}\left( \frac{\Phi \left( x\right) }{\lambda _{S_{i}E_{k}}y%
}\right) ^{h+1}\int_{\mathcal{L}_{2}}\frac{{\small \Gamma }\left(
1+h-s\right) {\small \Gamma }\left( -s\right) }{{\small \Gamma }\left(
1-s\right) }  \notag \\
& \times {\small \Gamma }\left( m_{S_{J}P}-s,\varphi _{J}\right) {\small %
\Gamma }\left( {L_{E_{k}}m_{S_{J}E_{k}}}+s\right) \left( 
\frac{{\small \eta }}{y}\right) ^{-s}{\small ds}  \notag \\
& =\left( \frac{\Phi \left( x\right) }{\lambda _{S_{i}E_{k}}y}\right)
^{h+1}\Theta _{2}^{(h)}\left( y\right) ,  
\end{align}%
where $\ \Theta _{2}^{(h)}\left( y\right) =G_{3,2}^{1,3}\left( \frac{\eta }{y%
}\left\vert 
\begin{array}{c}
\left( \mathcal{\zeta }_{J},\varphi _{J}\right) ,(1,0),(-h,0);- \\ 
({L_{E_{k}}m_{S_{J}E_{k}}},0);(0,0)%
\end{array}%
\right. \right) ,$ ${\small \eta =}\frac{\varsigma _{i}^{(k,J)}\Phi \left(
x\right) }{\overline{{\small \gamma }}_{I}},\mathcal{\zeta }%
_{J}=1-m_{S_{J}P},$ and $\varsigma _{i}^{(k,J)}$ is being defined in %
{Theorem 1}.

Finally, the conditional CDF of $\gamma _{1E}^{(i,k,J)}$ can be expressed by
substituting (\ref{T1_exp_final}) and (\ref{T2}) into (\ref{Fgamma1E_exp2})
and then replacing it into (\ref{Fgamma1Eb}), yields

\begin{align}
F_{\gamma _{1E}^{(i,k,J)}|g_{S_{i}P}=x}(y)& =1-{\small \Psi }%
_{i}^{(k)}\left( y\right) {\small \beta }_{k}^{(J)}
\label{Fgamma1E_exp_final} \notag\\
& \times \sum_{h{\small =0}}^{{L_{E_{k}}m_{S_{i}E_{k}}}{%
\small -1}}\frac{\binom{{L_{E_{k}}m_{S_{i}E_{k}}}{\small -1}}{%
{\small h}}\Phi ^{h+1}\left( x\right) }{\left( \lambda _{S_{i}E_{k}}y\right)
^{h+1}}   \notag \\
& \times \left[ 
\begin{array}{c}
\gamma \left( m_{S_{J}P},\varphi _{J}\right) \Theta _{1}^{(h)}\left( y\right)
\\ 
+\Theta _{2}^{(h)}\left( y\right)%
\end{array}%
\right] . 
\end{align}

Now, the remaining last previous step in this proof consists of computing $%
\Xi _{1}^{(i,k,J)}\left( x\right) .$ Indeed, by differentiating (\ref%
{F_gammaR}) and using (\ref{f_gq}) alongside with (\ref{Fgamma1E_exp_final}%
), (\ref{E1_exp1}) can be rewritten for a positive integer $%
{L_{R}m_{S_{i}R}}$ as 
\begin{align}
\Xi _{1}^{(i,k,J)}\left( x\right) & =\frac{\Gamma \left( %
{L_{R}m_{S_{i}R}},\frac{\xi _{S_{i}R}}{\Phi \left( x\right) }%
\right) }{\gamma \Gamma ({L_{R}m_{S_{i}R})}}  \label{E1_exp2}
\notag\\
& -\alpha _{i}^{(k,J)}\sum_{h{\small =0}}^{%
{L_{E_{k}}m_{S_{i}E_{k}}}{\small -1}}\frac{\Omega
_{h}^{(i,k)}e^{-\frac{\xi _{S_{i}R}}{\Phi \left( x\right) }}}{\left( \Phi
\left( x\right) \right) ^{{L_{E_{k}}m_{S_{i}E_{k}}}+%
{L_{R}m_{S_{i}R}}-h-2}}  \notag \\
& \times \sum_{l=0}^{{L_{R}m_{S_{i}R}}-1}\Upsilon _{l}^{(i)}%
\left[ \gamma \left( m_{S_{J}P},\varphi _{J}\right) \mathcal{U}_{1}^{(h,l)}+%
\mathcal{U}_{2}^{(h,l)}\right] ,  
\end{align}%
where $\xi _{S_{i}R},$ $\Omega _{h}^{(i,k)},$ $\alpha _{i}^{(k,J)},$ and $%
\Upsilon _{l}^{(i)}$ are defined in (\ref{Xi}), (\ref{Omega_h}), (\ref%
{Alpha_ikJ}), (\ref{Y_l}), respectively, and 
\begin{equation}
\mathcal{U}_{a}^{(h,l)}=\int_{0}^{\infty }y^{%
{L_{E_{k}}m_{S_{i}E_{k}}}+l-h-1}e^{-\frac{\varpi _{i}^{(k)}}{%
\Phi \left( x\right) }y}\Theta _{a}^{(h)}\left( y\right) dy,\text{ }%
a=\{1,2\},  \label{U1_exp1}
\end{equation}%
with $\varpi _{i}^{(k)}$ is being defined in {Theorem 1}.

The two above terms can be expressed as 
\begin{equation}
\mathcal{U}_{1}^{(h,l)}=\left( \frac{\Phi \left( x\right) }{\varpi _{i}^{(k)}%
}\right) ^{l+{L_{E_{k}}m_{S_{i}E_{k}}}-h}\mathcal{M}%
_{1}^{(h,l)}\left( \varpi _{i}^{(k)}\theta _{i}^{(k,J)}\right) ,
\label{U1_final}
\end{equation}%
\begin{align}
\mathcal{U}_{2}^{(h,l)}& =\frac{1}{2\pi j}\int_{\mathcal{L}_{3}}\frac{\Gamma
\left({L_{E_{k}}m_{S_{i}E_{k}}}+l-h+s\right) {\small \Gamma }%
\left( m_{S_{J}P}-s,\varphi _{J}\right) }{\left( \frac{\Phi \left( x\right) 
}{\varpi _{i}^{(k)}}\right) ^{-{L_{E_{k}}m_{S_{i}E_{k}}}-l+h}%
{\small \Gamma }\left( 1-s\right) }  \notag \\
& \times \frac{{\small \Gamma }\left( %
{L_{E_{k}}m_{S_{J}E_{k}}}+s\right) {\small \Gamma }\left(
-s\right) {\small \Gamma }\left( 1+h-s\right) }{\left( \frac{\varsigma
_{i}^{(k,J)}\varpi _{i}^{(k)}}{\overline{{\small \gamma }}_{I}}\right) ^{s}}%
{\small ds},  \notag \\
& =\left( \frac{\Phi \left( x\right) }{\varpi _{i}^{(k)}}\right) ^{%
{L_{E_{k}}m_{S_{i}E_{k}}}+l-h}\mathcal{M}_{2}^{(h,l)}\left( 
\frac{\varsigma _{i}^{(k,J)}\varpi _{i}^{(k)}}{\overline{{\small \gamma }}%
_{I}}\right) ,  \label{U2_final}
\end{align}%
where $\mathcal{M}_{1}^{(h,l)}\left( \bullet \right) $ and $\mathcal{M}%
_{2}^{(h,l)}\left( \bullet \right) $ are defined in (\ref{M1}) and (\ref{M2}%
), respectively. Note that (\ref{U1_final}) follows relying on \cite[Eq.
(07.34.21.0088.01)]{Wolfram}.

Henceforth, substituting (\ref{U1_final}) and (\ref{U2_final}) into (\ref%
{E1_exp2}), yields 
\begin{align}
\Xi _{1}^{(i,k,J)}\left( x\right) & =\frac{\Gamma \left( %
{L_{R}m_{S_{i}R}},\frac{\xi _{S_{i}R}}{\Phi \left( x\right) }%
\right) }{\gamma \Gamma ({L_{R}m_{S_{i}R}})}
\label{E1_finall}  \notag\\
& -\alpha _{i}^{(k,J)}\sum_{h{\small =0}}^{%
{L_{E_{k}}m_{S_{i}E_{k}}}{\small -1}}\frac{\Omega
_{h}^{(i,k)}e^{-\frac{\xi _{S_{i}R}}{\Phi \left( x\right) }}}{\left( \Phi
\left( x\right) \right) ^{{L_{E_{k}}m_{S_{i}E_{k}}}+%
{L_{R}m_{S_{i}R}}-h-2}}  \notag \\
& \times \sum_{l=0}^{{L_{R}m_{S_{i}R}}-1}\Upsilon
_{l}^{(i)}\left( \frac{\Phi \left( x\right) }{\varpi _{i}^{(k)}}\right) ^{%
{L_{E_{k}}m_{S_{i}E_{k}}}+l-h}  \notag \\
& \times \left[ 
\begin{array}{c}
\gamma \left( m_{S_{J}P},\varphi _{J}\right) \mathcal{M}_{1}^{(h,l)}\left(
\varpi _{i}^{(k)}\theta _{i}^{(k,J)}\right) \\ 
+\mathcal{M}_{2}^{(h,l)}\left( \frac{\varsigma _{i}^{(k,J)}\varpi _{i}^{(k)}%
}{\overline{{\small \gamma }}_{P}}\right)%
\end{array}%
\right] . 
\end{align}%
Now, replacing (\ref{E1_finall}) into (\ref{SOP1_exp2}), we obtain

\begin{align}
SOP_{1}^{(i,k,J)}& =1-\frac{\Lambda _{1}}{\Gamma (%
{L_{R}m_{S_{i}R}})}+\gamma \alpha _{i}^{(k,J)}
\label{SOP1_exp5} \notag\\
& \times \sum_{h{\small =0}}^{{L_{E_{k}}m_{S_{i}E_{k}}}{%
\small -1}}\Omega _{h}^{(i,k)}\sum_{l=0}^{{L_{R}m_{S_{i}R}}-1}%
\frac{\Upsilon _{l}^{(i)}\Lambda _{2}}{\left( \varpi _{i}^{(k)}\right) ^{l+%
{L_{E_{k}}m_{S_{i}E_{k}}}-h}}  \notag \\
& \times \left[ 
\begin{array}{c}
\gamma \left( m_{S_{J}P},\varphi _{J}\right) \mathcal{M}_{1}^{(h,l)}\left(
\varpi _{i}^{(k)}\theta _{i}^{(k,J)}\right) \\ 
+\mathcal{M}_{2}^{(h,l)}\left( \frac{\varsigma _{i}^{(k,J)}\varpi _{i}^{(k)}%
}{\overline{{\small \gamma }}_{I}}\right)%
\end{array}%
\right] ,  
\end{align}%
where 
\begin{align}
& \Lambda _{1}=\int_{0}^{\infty }f_{g_{S_{i}P}}(x)\Gamma \left( %
{L_{R}m_{S_{i}R}},\frac{\xi _{S_{i}R}}{\Phi \left( x\right) }%
\right) dx  \label{A1_final}  \notag\\
& \quad \overset{(a)}{=}\frac{1}{\Gamma \left( m_{S_{i}P}\right) }\left[ 
\begin{array}{c}
\gamma \left( m_{S_{i}P},\varphi _{S_{i}}\right) \Gamma \left( %
{L_{R}m_{S_{i}R}},\frac{\xi _{S_{i}R}}{\overline{{\small %
\gamma }}_{S_{i}}}\right) \\ 
+\mathcal{M}_{3}\left( \frac{\xi _{S_{i}R}}{\lambda _{S_{i}P}\overline{%
\gamma }_{P}}\right)%
\end{array}%
\right] , 
\end{align}%
\begin{align}
\Lambda _{2}& =\int_{0}^{\infty }\frac{f_{g_{S_{i}P}}(x)e^{-\frac{\xi
_{S_{i}R}}{\Phi \left( x\right) }}}{\left( \Phi \left( x\right) \right) ^{%
{L_{R}m_{S_{i}R}}-l-1}}dx  \label{A2_exp2}  \notag\\
& =\frac{\gamma \left( m_{S_{i}P},\varphi _{S_{i}}\right) e^{-\frac{\xi
_{S_{i}R}}{\overline{{\small \gamma }}_{S_{i}}}}}{\Gamma \left(
m_{S_{i}P}\right) \overline{{\small \gamma }}_{S_{i}}^{%
{L_{R}m_{S_{i}R}}-l-1}}+\frac{\lambda _{S_{i}P}^{m_{S_{i}P}}}{%
\Gamma \left( m_{S_{i}P}\right) \overline{{\small \gamma }}_{I}^{%
{L_{R}m_{S_{i}R}}-l-1}}  \notag \\
& \times \frac{\Gamma \left( m_{S_{i}P}+{L_{R}m_{S_{i}R}}%
-l-1,\varphi _{S_{i}}+\frac{\xi _{S_{i}R}}{\overline{\gamma }_{S_{i}}}%
\right) }{\left( \lambda _{S_{i}P}+\frac{\xi _{S_{i}R}}{\overline{\gamma }%
_{P}}\right) ^{m_{S_{i}P}+{L_{R}m_{S_{i}R}}-l-1}}, 
\end{align}%
with $\mathcal{M}_{3}\left( \bullet \right) $ is defined in (\ref{M3}).
Equality $(a)$ holds by replacing $\Phi \left( x\right) $ by their values
and along using \cite[Eqs. (06.06.26.0005.01), (07.34.21.0088.01)]{Wolfram}.

By substituting (\ref{A1_final}) and (\ref{A2_exp2}) into (\ref{SOP1_exp5}),
(\ref{SOP_1k_final}) is attained.

\subsection{Expression of $SOP$ at the Second Hop}

In like manner to $SOP_{1}^{(i,k)},$ $SOP_{2}^{(k)}$ can be expressed as 
\begin{equation}
SOP_{2}^{(k)}=1-\gamma \int_{0}^{\infty }f_{g_{RP}}(x)\Xi _{3}^{(k)}\left(
x\right) dx,  \label{sop2}
\end{equation}
with 
\begin{equation}
\Xi _{3}^{(k)}\left( x\right) =\int_{0}^{\infty }f_{\gamma
_{D}|g_{RP}=x}\left( \gamma +\gamma y-1\right) F_{\gamma
_{2E}^{(k)}|g_{RP}=x}\left( y\right) dy.  \label{E2_exp1}
\end{equation}
One can notice from (\ref{E2_exp1}) that in order to calculate $%
SOP_{2}^{(k)},$ it is necessary to find first the conditional CDFs of $%
\gamma _{D}$ and $\gamma _{2E}^{(k)}$ for a given $g_{RP}.$

\begin{itemize}
\item Conditional CDFs of $\gamma _{D}$ and $\gamma _{2E}^{(k)}$
\end{itemize}

Let's define $Y_{RD}=\sum_{t=1}^{L}g_{RD_{t}}$. In a similar manner to (\ref%
{F_gammaR}), the conditional CDFs of $\gamma _{D}$ and $\gamma _{2E}^{(k)}$
are given, respectively, by 
\begin{equation}
F_{\gamma _{D}|g_{RP}=x}(z)=F_{Y_{RD}}\left( \frac{z}{\mathcal{D}\left(
x\right) }\right) ,  \label{F_gammaD_exp1}
\end{equation}%
\begin{equation}
F_{\gamma _{2E}^{(k)}|g_{RP}=x}\left( y\right) ={F_{%
\mathcal{E}_{RE_{k}}}}\left( \frac{y}{\mathcal{D}(x)}\right) ,
\label{F_2E_exp1}
\end{equation}%
where $\mathcal{E}_{RE_{k}}=\sum_{u=1}^{L_{E_{k}}}g_{RE_{k}^{(u)}},$ $%
\mathcal{D}\left( x\right) =$ $\overline{{\small \gamma }}_{R}$ for $x\leq 
\overline{{\small \gamma }}_{I}/\overline{{\small \gamma }}_{R}$ and $%
\mathcal{D}\left( x\right) =$ $\overline{{\small \gamma }}_{I}/x$ for $x>%
\overline{{\small \gamma }}_{I}/\overline{\gamma }_{R}.$ 
{It follows, in a
similar manner to $\mathcal{E}_{S_{i}R},$ that $\mathcal{E}_{RE_{k}}$ is also
Gamma distributed with parameters {$L_{E_{k}}m_{RE_{k}}$} and $\lambda
_{RE_{k}}.$}

\begin{itemize}
\item Expression of $\Xi _{3}^{(k)}\left( x\right) $
\end{itemize}

It is worthwhile that $Y_{RD}$ is Gamma distributed for i.i.d Nakagami-$m$
random variables with shape and scale parameters $L_{D}m_{RD}$ and $\lambda
_{RD}$, respectively. That is 
\begin{align}
& \Xi _{3}^{(k)}\left( x\right) \overset{(a)}{=}\frac{\lambda
_{RD}^{L_{D}m_{RD}}e^{-\frac{\xi _{RD}}{\mathcal{D}\left( x\right) }}}{%
\Gamma (L_{D}m_{RD})\Gamma \left( {L_{E_{k}}m_{RE_{k}}}%
\right) \left( \mathcal{D}\left( x\right) \right) ^{L_{D}m_{RD}-1}}
\label{E2_final} \notag\\
& \quad \quad \times \sum_{j=0}^{L_{D}m_{RD}-1}\mathcal{B}_{j}\gamma
^{j}\int_{0}^{\infty }y^{j}e^{-\frac{\lambda _{RD}\gamma }{\mathcal{D}\left(
x\right) }y}\gamma \left( {L_{E_{k}}m_{RE_{k}}},\frac{\lambda
_{_{RE_{k}}}}{\mathcal{D}(x)}y\right)  \notag \\
& \quad \quad \overset{(b)}{=}\frac{\lambda _{RD}^{L_{D}m_{RD}}e^{-\frac{\xi
_{RD}}{\mathcal{D}\left( x\right) }}}{\gamma \Gamma (L_{D}m_{RD})\Gamma
\left( {L_{E_{k}}m_{RE_{k}}}\right) \mathcal{D}%
^{L_{D}m_{RD}-1}\left( x\right) }
\notag \\
& \quad \quad \times \sum_{j=0}^{L_{D}m_{RD}-1}\mathcal{B}_{j} \left( \frac{\mathcal{D}\left( x\right) }{\lambda _{RD}}%
\right) ^{j+1}G_{2,2}^{1,2}\left( \frac{\lambda _{_{RE_{k}}}}{\lambda
_{RD}\gamma }\left\vert 
\begin{array}{c}
-j,1;- \\ 
{L_{E_{k}}m_{RE_{k}}};0%
\end{array}%
\right. \right) ,  
\end{align}%
where $\mathcal{B}_{j}$ is defined in (\ref{Bj}). Note that step $(a)$ holds
by substituting (\ref{F_gammaD_exp1}) and (\ref{F_2E_exp1}) into (\ref%
{E2_exp1}), while equality $(b)$ follows by using \cite[Eqs.
(06.06.26.0004.01), (07.34.21.0088.01)]{Wolfram}.

Substituting (\ref{E2_final}) into (\ref{sop2}), yields 
\begin{align}
SOP_{2}^{(k)}& =1-\frac{\lambda _{RD}^{L_{D}m_{RD}}}{\Gamma \left( %
{L_{E_{k}}m_{RE_{k}}}\right) \Gamma (L_{D}m_{RD})}%
\sum_{j=0}^{L_{D}m_{RD}-1}\frac{\mathcal{B}_{j}}{\lambda _{RD}^{j+1}}
\label{SOP2_exp4} \notag\\
& \times \mathcal{J}_{j}G_{2,2}^{1,2}\left( \frac{\lambda _{RE_{k}}}{\lambda
_{RD}\gamma }\left\vert 
\begin{array}{c}
-j,1;- \\ 
{L_{E_{k}}m_{RE_{k}}};0%
\end{array}%
\right. \right) ,  
\end{align}%
where 
\begin{align}
& \mathcal{J}_{j}\mathcal{=}\int_{0}^{\infty }f_{g_{RP}}(x)\left( \mathcal{D}%
\left( x\right) \right) ^{-L_{D}m_{RD}+j+1}e^{-\frac{\xi _{RD}}{\mathcal{D}%
\left( x\right) }}dx  \label{J_exp2} \notag\\
& \quad \overset{(a)}{=}\frac{1}{\Gamma (m_{RP})}\left[ 
\begin{array}{c}
\frac{e^{-\frac{\xi _{RD}}{\overline{{\small \gamma }}_{R}}}\gamma \left(
m_{RP},\varphi _{R}\right) }{\overline{{\small \gamma }}%
_{R}^{L_{D}m_{RD}-j-1}} \\ 
+\frac{\lambda _{RP}^{m_{RP}}\Gamma \left( \upsilon _{j},\varphi _{R}+\frac{%
\xi _{RD}}{\overline{{\small \gamma }}_{R}}\right) }{\overline{{\small %
\gamma }}_{I}^{L_{D}m_{RD}-j-1}\left( \frac{\xi _{RD}}{\overline{{\small %
\gamma }}_{I}}+\lambda _{RP}\right) ^{\upsilon _{j}}}%
\end{array}%
\right] ,  
\end{align}%
where $\upsilon _{j}=$ $L_{D}m_{RD}+m_{RP}-j-1,$ $\varphi _{R}$ is defined
in {Theorem 1}. Here step $(a)$ is obtained by replacing $%
\mathcal{D}\left( x\right) $ by its values and using (\ref{f_gq}) alongside
with Eqs. (3.381.1) and (3.381.3) of \cite{table}.

By considering $\frac{\overline{{\small \gamma }}_{I}}{\overline{{\small %
\gamma }}_{R}}=\delta $ and substituting (\ref{J_exp2}) into\ (\ref%
{SOP2_exp4}), one can obtain (\ref{SOP2_final}) which concludes the proof.

\section*{Appendix B: proof of theorem 2}
\addtocounter{subsection}{-2}
In this section, we make use of the residues theorem in order to find the approximate expressions of Meijer-G's function given in (\ref{SOP_1k_final}).

\subsection{Asymptotic Expression of $SOP_{1}^{(i,k,J)}$}

\subsubsection{Case 1: Presence of Jammer}

The Meijer-G's functions $\mathcal{M}_{1}^{(h,l)}\left( z\right) $ and $%
\mathcal{M}_{2}^{(h,l)}\left( z\right) $ given in (\ref{SOP_1k_final}) can
be expressed in terms of complex integral as

\begin{align}
\mathcal{M}_{1}^{(h,l)}\left( z\right) & =\frac{1}{2\pi j}\int_{\mathcal{L}%
_{3}}\frac{\Gamma \left( {L_{E_{k}}m_{S_{J}E_{k}}}+s\right)
\Gamma \left( 1+h-s\right) }{\Gamma \left( 1-s\right) }  \notag \\
& \times \Gamma \left( {L_{E_{k}}m_{S_{i}E_{k}}}+l-h+s\right)
\Gamma \left( -s\right) z^{-s}ds,  
  \label{M1_integral}
\end{align}%
and%
\begin{align}
\mathcal{M}_{2}^{(h,l)}\left( z\right) & =\frac{1}{2\pi j}\int_{\mathcal{L}%
_{3}}\frac{{\small \Gamma }\left({L_{E_{k}}m_{S_{J}E_{k}}}%
+s\right) {\small \Gamma }\left( 1+h-s\right) }{{\small \Gamma }\left(
1-s\right) }  \notag \\
& \times \Gamma \left( {L_{E_{k}}m_{S_{i}E_{k}}}+l-h+s\right) 
{\small \Gamma }\left( -s\right)  \notag \\
& \times {\small \Gamma }\left( m_{S_{J}P}-s,\varphi _{J}\right) z^{-s}%
{\small ds} {.}  \label{M2_integral}
\end{align}%
It is noteworthy that the conditions of \cite[Theorem 1.5]{Kilbas2004} are
satisfied. That is, the two above functions can be written as an infinite
sum of the poles belonging to the left half plan of $\mathcal{L}_{3}$.
Furthermore, as the upper incomplete gamma function in (\ref{M2_integral})
is always finite for $\varphi _{J}\neq 0,$ it follows that the integrand
functions of the two above equations have the same poles. Additionally, it
is clearly seen that the order of the left poles depends on the values of $%
{L_{E_{k}}m_{S_{J}E_{k}}},$ $%
{L_{E_{k}}m_{S_{i}E_{k}}},$ $h,$ and $l$. Owing to this fact,
three cases can be distinguished{:}

\begin{itemize}
\item $-{L_{E_{k}}m_{S_{J}E_{k}}}<-%
{L_{E_{k}}m_{S_{i}E_{k}}}-l+h$: In this case, the two
integrand functions given in (\ref{M1_integral}) and (\ref{M2_integral})
admit $-\chi _{h,l,r}$ with $\chi _{h,l,r}=%
{L_{E_{k}}m_{S_{i}E_{k}}}+l-h+r$ and\ $0\leq r\leq %
{L_{E_{k}}m_{S_{J}E_{k}}}-%
{L_{E_{k}}m_{S_{i}E_{k}}}-l+h-1$ as simple poles and $%
-\varrho _{r}$ with $\varrho _{r}={L_{E_{k}}m_{S_{J}E_{k}}}+r$
and $r$ natural number as poles of second-order.

\item $-{L_{E_{k}}m_{S_{J}E_{k}}}>-%
{L_{E_{k}}m_{S_{i}E_{k}}}-l+h:$ Under this condition, the
aforementioned integrands have $-\varrho _{r}$ with $0\leq r\leq %
{L_{E_{k}}m_{S_{i}E_{k}}}-%
{L_{E_{k}}m_{S_{J}E_{k}}}+l-h-1$ as simple poles and $-\chi
_{h,l,r}$ where $r\in 
\mathbb{N}
$ as poles of second-order.

\item $-{L_{E_{k}}m_{S_{J}E_{k}}}%
=-L_{E_{k}}m_{S_{i}E_{k}}-l+h:$ Under this assumption, the two integrands
admit only poles of second-order at $-\varrho _{r},$ $r\in 
\mathbb{N}
.$
\end{itemize}

$\mathbf{a.}$ $\mathbf{-}{L_{E_{k}}m_{S_{J}E_{k}}}\mathbf{<-}%
{L_{E_{k}}m_{S_{i}E_{k}}}\mathbf{-l+h}$

Relied on \cite[Theorem 1.5]{Kilbas2004}, $\mathcal{M}_{1}^{(h,l)}\left(
z\right) \ $can be rewritten as series of residues at the aforementioned
poles

\begin{align}
\mathcal{M}_{1}^{(h,l)}\left( z\right) & =\sum_{r=0}^{\varrho _{r}-\chi
_{h,l,r}-1}\lim_{s\rightarrow -\chi _{h,l,r}}Q_{1}\left( s,z\right)
\label{M1_residue_exp1}  \notag\\
& +\sum_{r=0}^{\infty }\lim_{s\rightarrow -\varrho _{r}}\frac{\partial
Q_{2}\left( s,z\right) }{\partial s}, 
\end{align}%
where 
\begin{align}
Q_{1}\left( s,z\right) & =\left( \chi _{h,l,r}+s\right) \Gamma \left( \chi
_{h,l,r}-r+s\right)  \label{Q1} \notag\\
& \times \frac{{\small \Gamma }\left( %
{L_{E_{k}}m_{S_{J}E_{k}}}+s\right) {\small \Gamma }\left(
-s\right) {\small \Gamma }\left( 1+h-s\right) }{{\small \Gamma }\left(
1-s\right) }z^{-s},  
\end{align}%
and 
\begin{align}
Q_{2}\left( s,z\right) & =(\varrho _{r}+s)^{2}\Gamma \left( %
{L_{E_{k}}m_{S_{i}E_{k}}}+l-h+s\right)  \notag \\
& \times \frac{{\small \Gamma }\left( %
{L_{E_{k}}m_{S_{J}E_{k}}}+s\right) {\small \Gamma }\left(
-s\right) {\small \Gamma }\left( 1+h-s\right) }{{\small \Gamma }\left(
1-s\right) }z^{-s}.  \notag \\
&  \label{Q2}
\end{align}%
Obviously, the limit of $Q_{1}\left( s,z\right) $ can be expressed as

\begin{align}
\lim_{s\rightarrow -\chi _{h,l,r}}Q_{1}\left( s,z\right) & =\frac{\left(
-1\right) ^{r}{\small \Gamma }\left( {L_{E_{k}}m_{S_{i}E_{k}}}%
+l+r+1\right) }{r!\chi _{h,l,r}}  \label{limit_Q1}  \notag\\
& \times {\small \Gamma }\left( {L_{E_{k}}m_{S_{J}E_{k}}}%
-\chi _{h,l,r}\right) z^{\chi _{h,l,r}}. 
\end{align}%
On the other hand, using 
{\cite[Eqs. (06.14.06.0026.01) and
(06.14.16.0003.01)]{Wolfram}} the partial derivative of $Q_{2}\left(
s,z\right) $ is given by 
\begin{align}
\frac{\partial Q_{2}\left( s,z\right) }{\partial s}& =\frac{(s+\varrho
_{r})^{2}\Gamma \left( \chi _{h,l,r}-r+s\right) {\small \Gamma }\left(
-s\right) }{{\small \Gamma }\left( 1-s\right) z^{s}}  \label{derivative_Q2}
 \notag\\
& \times {\small \Gamma }\left( {L_{E_{k}}m_{S_{J}E_{k}}}%
+s\right) {\small \Gamma }\left( 1+h-s\right) \mathcal{G}^{(h,l,r)}\left(
z,s\right) , 
\end{align}%
where $\mathcal{G}^{(h,l,r)}\left( z,s\right) =-\log z+\psi (r+1)+\psi
(\varrho _{r}-\chi _{h,l,r}+r+1)-\frac{1}{s}-\psi (1+h-s).$

{Replacing (\ref{derivative_Q2}) and} (\ref{limit_Q1}) %
{into} (\ref{M1_residue_exp1}), yields 
\begin{align}
\mathcal{M}_{1}^{(h,l)}\left( z\right) & =\sum_{r=0}^{\varrho _{r}-\chi
_{h,l,r}-1}\frac{\left( -1\right) ^{r}{\small \Gamma }\left( %
{L_{E_{k}}m_{S_{J}E_{k}}}-\chi _{h,l,r}\right) }{r!\chi
_{h,l,r}}  \notag \\
& \times {\small \Gamma }\left( {L_{E_{k}}m_{S_{i}E_{k}}}%
+l+r+1\right) z^{\chi _{h,l,r}}  \notag \\
& +\sum_{r=0}^{\infty }\frac{\left( -1\right) ^{\varrho _{r}-\chi _{h,l,r}}%
{\small \Gamma }\left( h+\varrho _{r}+1\right) }{\left( -%
{L_{E_{k}}m_{S_{i}E_{k}}}+\varrho _{r}-l+h\right) !k!\varrho
_{r}}  \notag \\
& \times z^{\varrho _{r}}\mathcal{G}^{(h,l,r)}\left( z,-\varrho _{r}\right) .
\label{M1_residue_final}
\end{align}

In similar manner$,$ $\mathcal{M}_{2}^{(h,l)}\left( z\right) $ can be
expressed as%
\begin{align}
\mathcal{M}_{2}^{(h,l)}\left( z\right) & =\sum_{k=0}^{\varrho _{r}-\chi
_{h,l,r}-1}\frac{\left( -1\right) ^{r}{\small \Gamma }\left( %
{L_{E_{k}}m_{S_{i}E_{k}}}+l+r+1\right) }{r!\chi
_{h,l,r}z^{-\chi _{h,l,r}}}  \notag \\
& \times {\small \Gamma }\left( m_{S_{J}P}+\chi _{h,l,r},\varphi _{J}\right) 
{\small \Gamma }\left( {L_{E_{k}}m_{S_{J}E_{k}}}-\chi
_{h,l,r}\right)  \notag \\
& +\sum_{r=0}^{\infty }\frac{\left( -1\right) ^{\varrho _{r}-\chi _{h,l,r}}%
{\small \Gamma }\left( 1+h+\varrho _{r}\right) }{\left( -%
{L_{E_{k}}m_{S_{i}E_{k}}}+\varrho _{r}-l+h\right) !r!\varrho
_{r}}z^{\varrho _{r}}  \notag \\
& \times \left\{ 
\begin{array}{c}
\left[ \mathcal{G}^{(h,l,r)}\left( z,-\varrho _{r}\right) -\log \left(
\varphi _{J}\right) \right] \\ 
\times {\small \Gamma }\left( m_{S_{J}P}+\varrho _{r},\varphi _{J}\right) -%
\mathcal{V}^{(r)}\left( \varrho _{r}\right)%
\end{array}%
\right\} ,  \notag \\
&  \label{M2_residue_Final}
\end{align}%
where $\mathcal{V}^{(r)}\left( \varrho _{r}\right) \mathcal{=}%
G_{2,3}^{3,0}\left( \varphi _{J}\left\vert 
\begin{array}{c}
-;1,1 \\ 
0,0,m_{S_{J}P}+\varrho _{r};-%
\end{array}%
\right. \right) .$

Interestingly, one can notice that $\varrho _{r}>\chi _{h,l,r}$ for $%
{L_{E_{k}}m_{S_{J}E_{k}}}>%
{L_{E_{k}}m_{S_{i}E_{k}}}+l-h.$ Consequently, the second
summation in the two above expressions can be ignored as $z$ approaches $0,$
i.e., 
\begin{align}
\mathcal{M}_{1}^{(h,l)}\left( z\right) & \sim \frac{{\small \Gamma }\left( %
{L_{E_{k}}(m_{S_{J}E_{k}}-m_{S_{i}E_{k}})}-l+h\right) }{%
\left( {L_{E_{k}}m_{S_{i}E_{k}}}+l-h\right) },
\label{M1_approximate_case1} \notag\\
& \times {\small \Gamma }\left( {L_{E_{k}}m_{S_{i}E_{k}}}%
+l+1\right) z^{{L_{E_{k}}m_{S_{i}E_{k}}}+l-h}  
\end{align}%
and 
\begin{align}
\mathcal{M}_{2}^{(h,l)}\left( z\right) & \sim \frac{{\small \Gamma }\left( %
{L_{E_{k}}m_{S_{i}E_{k}}}+l+1\right) z^{%
{L_{E_{k}}m_{S_{i}E_{k}}}+l-h}}{\left( %
{L_{E_{k}}m_{S_{i}E_{k}}}+l-h\right) }  \notag \\
& \times {\small \Gamma }\left( %
{L_{E_{k}}(m_{S_{J}E_{k}}-m_{S_{i}E_{k}})}-l+h\right)  \notag
\\
& \times {\small \Gamma }\left( m_{S_{J}P}+%
{L_{E_{k}}m_{S_{i}E_{k}}}+l-h,\varphi _{J}\right) .
\label{M2_approximate_case1}
\end{align}%
$\mathbf{b.}$\textbf{\ }$\mathbf{-}{L_{E_{k}}m_{S_{J}E_{k}}}%
\mathbf{>-}{L_{E_{k}}m_{S_{i}E_{k}}}\mathbf{-l+h}$

Analogously to the previous case, the integrals (\ref{M1_integral}) and (\ref%
{M2_integral}) can be computed relied on \cite[Theorem 1.5]{Kilbas2004} and 
\cite[Eq. (06.14.16.0003.01)]{Wolfram}, respectively, as 
\begin{align}
\mathcal{M}_{1}^{(h,l)}\left( z\right) & =\sum_{r=0}^{\chi _{h,l,r}-\varrho
_{r}-1}\frac{\Gamma \left({L_{E_{k}}m_{S_{i}E_{k}}}%
+l-h-\varrho _{r}\right) {\small \Gamma }\left( 1+h+\varrho _{r}\right) }{%
r!\varrho _{r}\left( -1\right) ^{-r}z^{-\varrho _{r}}}  \notag \\
& +\sum_{r=0}^{\infty }\frac{\left( -1\right) ^{\chi _{h,l,r}-\varrho _{r}}%
{\small \Gamma }\left( 1+h+\chi _{h,l,r}\right) }{\left( \chi _{h,l,r}-%
{L_{E_{k}}m_{S_{J}E_{k}}}\right) !r!\chi _{h,l,r}z^{-\chi
_{h,l,r}}}  \notag \\
& \times \left\{ 
\begin{array}{c}
\psi (r+1)+\psi (-{L_{E_{k}}m_{S_{J}E_{k}}}+\chi _{h,l,r}+1)
\\ 
-\psi (1+{L_{E_{k}}m_{S_{i}E_{k}}}+l+r)+\frac{1}{\chi _{h,l,r}%
}-\log z%
\end{array}%
\right\} ,  \notag \\
&  \label{M1_case2}
\end{align}%
and 
\begin{align}
\mathcal{M}_{2}^{(h,l)}\left( z\right) & =\sum_{r=0}^{\chi _{h,l,r}-\varrho
_{r}-1}\frac{\left( -1\right) ^{r}\Gamma \left( %
{L_{E_{k}}m_{S_{i}E_{k}}}+l-h-\varrho _{r}\right) }{r!\varrho
_{r}}  \notag \\
& \times {\small \Gamma }\left( 1+h+\varrho _{r}\right) z^{\varrho _{r}}%
{\small \Gamma }\left( m_{S_{J}P}+\varrho _{r},\varphi _{J}\right)  \notag \\
& +\sum_{r=0}^{\infty }\frac{\left( -1\right) ^{\chi _{h,l,r}-\varrho _{r}}%
{\small \Gamma }\left( 1+h+\chi _{h,l,r}\right) }{\left( \chi
_{h,l,r}-\varrho _{r}\right) !r!\chi _{h,l,r}z^{-\chi _{h,l,r}}}  \notag \\
& \times \left[ \mathcal{Z}-\mathcal{V}^{(r)}\left( \chi _{h,l,r}\right) %
\right]  \label{M2_case2}
\end{align}%
where 
\begin{equation}
\mathcal{Z=}{\small \Gamma }\left( m_{S_{J}P}+\chi _{h,l,r},\varphi
_{J}\right) \left[ 
\begin{array}{c}
-\log z+\psi (r+1) \\ 
+\psi (-{L_{E_{k}}m_{S_{J}E_{k}}}+\chi _{h,l,r}+1) \\ 
-\psi ({L_{E_{k}}m_{S_{i}E_{k}}}+l-h+r) \\ 
-\psi ({L_{E_{k}}m_{S_{i}E_{k}}}+l+r+1) \\ 
+\psi (1+\chi _{h,l,r})-\log \left( \varphi _{J}\right)%
\end{array}%
\right] .
\end{equation}

One can notice that as $\varrho _{r}<\chi _{h,l,r},$ $\mathcal{M}%
_{1}^{(h,l)}\left( z\right) $ and $\mathcal{M}_{2}^{(h,l)}\left( z\right) $
can be approximated when $z$ tends to $0$ 
\begin{align}
\mathcal{M}_{1}^{(h,l)}\left( z\right) & \sim \frac{\Gamma \left( %
{L_{E_{k}}m_{S_{i}E_{k}}}-%
{L_{E_{k}}m_{S_{J}E_{k}}}+l-h\right) }{%
{L_{E_{k}}m_{S_{J}E_{k}}}}  \label{M1_approximate_case2} \notag\\
& \times {\small \Gamma }\left( {L_{E_{k}}m_{S_{J}E_{k}}}%
+h+1\right) z^{{L_{E_{k}}m_{S_{J}E_{k}}}}  
\end{align}%
and 
\begin{align}
\mathcal{M}_{2}^{(h,l)}\left( z\right) & \sim \frac{{\small \Gamma }\left( %
{L_{E_{k}}m_{S_{J}E_{k}}}+h+1\right) {\small \Gamma }\left(
m_{S_{J}P}+{L_{E_{k}}m_{S_{J}E_{k}}},\varphi _{J}\right) }{%
{L_{E_{k}}m_{S_{J}E_{k}}}}  \notag \\
& \times \Gamma \left( {L_{E_{k}}m_{S_{i}E_{k}}}-%
{L_{E_{k}}m_{S_{J}E_{k}}}+l-h\right) z^{%
{L_{E_{k}}m_{S_{J}E_{k}}}}.  \label{M2_approximate_case2}
\end{align}%
$\mathbf{c.}$ $\mathbf{-}{L_{E_{k}}m_{S_{J}E_{k}}}\mathbf{=-}%
{L_{E_{k}}m_{S_{i}E_{k}}}-l+h$

For this case, the two complex integrals given in (\ref{M1_integral}) and (%
\ref{M2_integral}) can be expressed by performing some algebraic operations
as 
\begin{align}
\mathcal{M}_{1}^{(h,l)}\left( z\right) & =\sum_{r=0}^{\infty }\frac{\left(
-1\right) ^{\varrho _{r}-\chi _{h,l,r}}{\small \Gamma }\left( 1+h+\varrho
_{r}\right) }{\left( -{L_{E_{k}}m_{S_{i}E_{k}}}+\varrho
_{r}-l+h\right) !k!\varrho _{r}}  \notag \\
& \times z^{\varrho _{r}}\mathcal{G}^{(h,l,r)}\left( z,-\varrho _{r}\right) .
\label{M1_case3}
\end{align}%
and%
\begin{align}
\mathcal{M}_{2}^{(h,l)}\left( z\right) & =\sum_{r=0}^{\infty }\frac{\left(
-1\right) ^{\varrho _{r}-\chi _{h,l,r}}{\small \Gamma }\left( 1+h+\varrho
_{r}\right) }{\left( \varrho _{r}-{L_{E_{k}}m_{S_{i}E_{k}}}%
-l+h\right) !r!\varrho _{r}}z^{\varrho _{r}}  \notag \\
& \times \left\{ 
\begin{array}{c}
\left[ \mathcal{G}^{(h,l,r)}\left( z\right) -\log \left( \varphi _{J}\right) %
\right] \\ 
\times {\small \Gamma }\left( m_{S_{J}P}+\varrho _{r},\varphi _{J}\right) -%
\mathcal{V}^{(r)}\left( \varrho _{r}\right)%
\end{array}%
\right\} .  \notag \\
&  \label{M2_case3}
\end{align}

Again, $\mathcal{M}_{1}^{(h,l)}\left( z\right) $ and $\mathcal{M}%
_{2}^{(h,l)}\left( z\right) $ can be approximated as $z$ approaches $0$ by 
\begin{align}
\mathcal{M}_{1}^{(h,l)}\left( z\right) & \sim \frac{\left( -1\right) ^{%
{L_{E_{k}}m_{S_{J}E_{k}}}-{L_{R}m_{S_{i}R}}%
+1}z^{{L_{E_{k}}m_{S_{J}E_{k}}}}\log z,}{%
{L_{E_{k}}m_{S_{J}E_{k}}}}  \notag \\
& \times {\small \Gamma }\left( {L_{E_{k}}m_{S_{i}E_{k}}}+%
{L_{E_{k}}m_{S_{J}E_{k}}}\right) 
\label{M1_approximate_case3}
\end{align}%
and 
\begin{align}
\mathcal{M}_{2}^{(h,l)}\left( z\right) & \sim \frac{\left( -1\right) ^{%
{L_{E_{k}}m_{S_{J}E_{k}}}-{L_{R}m_{S_{i}R}}+1}%
{\small \Gamma }\left( {L_{E_{k}}}\left(m_{S_{i}E_{k}}+%
m_{S_{J}E_{k}}\right)\right) }{{L_{E_{k}}m_{S_{J}E_{k}}}} 
\notag \\
& \times {\small \Gamma }\left( m_{S_{J}P}+%
{L_{E_{k}}m_{S_{J}E_{k}}},\varphi _{J}\right) z^{%
{L_{E_{k}}m_{S_{J}E_{k}}}}\log \left( z\right) .
\label{M2_approximate_case3}
\end{align}

Finally, the Meijer's G-function $\mathcal{M}_{3}\left( z\right) $ defined
in (\ref{M3}) can be written in terms of complex integral as 
\begin{equation}
\mathcal{M}_{3}\left( z\right) =\frac{1}{2\pi j}\int_{\mathcal{L}_{4}}\frac{%
{\small \Gamma }\left( {L_{R}m_{S_{i}R}}+s\right) {\small %
\Gamma }\left( m_{S_{i}P}-s,\varphi _{S_{i}}\right) }{{\small s}}z^{-s}%
{\small ds}.  \label{M3_integral}
\end{equation}%
It is worth mentioning that the conditions of \cite{Kilbas2004} are applied
also here. Thus, the above integrand function can be written as an infinite
sum of the left poles in $\mathcal{L}_{4}$. \ Besides, that integrand admits
only poles of the first order at $0$ and $-{L_{R}m_{S_{i}R}}%
-r,$ $r\in 
\mathbb{N}
.$ That is 
\begin{align}
\mathcal{M}_{3}\left( z\right) & ={\small \Gamma }\left( %
{L_{R}m_{S_{i}R}}\right) {\small \Gamma }\left(
m_{S_{i}P},\varphi _{S_{i}}\right)  \label{M3_approximate} \\
& +\sum_{r=0}^{\infty }\frac{\left( -1\right) ^{r+1}{\small \Gamma }\left(
m_{S_{i}P}+{L_{R}m_{S_{i}R}}+r,\varphi _{S_{i}}\right) }{%
r!\left( {L_{R}m_{S_{i}R}}+r\right) z^{-%
{L_{R}m_{S_{i}R}}-r}}  \notag \\
& \overset{(a)}{\sim }{\small \Gamma }\left( {L_{R}m_{S_{i}R}}%
\right) {\small \Gamma }\left( m_{S_{i}P},\varphi _{S_{i}}\right) -\frac{%
{\small \Gamma }\left( m_{S_{i}P}+{L_{R}m_{S_{i}R}},\varphi
_{S_{i}}\right) }{{L_{R}m_{S_{i}R}}z^{-%
{L_{R}m_{S_{i}R}}}},  \notag
\end{align}%
with step $(a)$ follows by considering only the first term of the infinite
summation when $z$ tends to $0.$

Finally, armed by \cite[Eq. (8.354.2)]{table} the upper incomplete Gamma
given in (\ref{SOP_1k_final}) can be approximated for small values of $z$ as

\begin{equation}
\Gamma \left( {L_{R}m_{S_{i}R}},\frac{\sigma _{i}\xi _{S_{i}R}%
}{\overline{{\small \gamma }}_{I}}\right) \sim \Gamma \left( %
{L_{R}m_{S_{i}R}}\right) -\frac{1}{%
{L_{R}m_{S_{i}R}}}\left( \frac{\sigma _{i}\xi _{S_{i}R}}{%
\overline{{\small \gamma }}_{I}}\right) ^{{L_{R}m_{S_{i}R}}}.
\label{gamma_approximate}
\end{equation}

Interestingly, the $SOP_{1}^{(i,k,J)}$ can finally be approximated in high
SNR regime (i.e., $\overline{{\small \gamma }}_{I}\rightarrow \infty )$ by
considering three cases, namely ${L_{R}m_{S_{i}R}}<%
{L_{E_{k}}m_{S_{J}E_{k}}},$ ${L_{R}m_{S_{i}R}}>%
{L_{E_{k}}m_{S_{J}E_{k}}},$ and $%
{L_{R}m_{S_{i}R}}={L_{E_{k}}m_{S_{J}E_{k}}}.$

\begin{itemize}
\item ${L_{R}m_{S_{i}R}}<%
{L_{E_{k}}m_{S_{J}E_{k}}}$
\end{itemize}

Substituting (\ref{M1_approximate_case1}), (\ref{M2_approximate_case1}), (%
\ref{M3_approximate}), and (\ref{gamma_approximate}) into (\ref{SOP_1k_final}%
), and by considering $h={L_{E_{k}}m_{S_{i}E_{k}}}{\small -1,}
$ $SOP_{1}^{(i,k,J)}$ can be approximated\ as 
\begin{equation}
SOP_{1}^{(i,k,J)}\sim \frac{\mathcal{C}_{1}^{(i,k,J)}}{\overline{{\small %
\gamma }}_{I}^{{L_{R}m_{S_{i}R}}}},
\label{SOP1_approximate_case1}
\end{equation}%
where $\mathcal{C}_{1}^{(i,k,J)}$ is given in (\ref{C1}).

\begin{itemize}
\item ${L_{R}m_{S_{i}R}}>%
{L_{E_{k}}m_{S_{J}E_{k}}}$
\end{itemize}

Incorporating (\ref{M1_approximate_case2}), (\ref{M2_approximate_case2}), (%
\ref{M3_approximate}), and (\ref{gamma_approximate}) into (\ref{SOP_1k_final}%
), and by considering $l={L_{R}m_{S_{i}R}}{\small -1,}$ $%
SOP_{1}^{(i,k,J)}$ can be approximated as 
\begin{equation}
SOP_{1}^{(i,k,J)}\sim \frac{\mathcal{C}_{2}^{(i,k,J)}}{\overline{{\small %
\gamma }}_{I}^{{L_{E_{k}}m_{S_{J}E_{k}}}}},
\label{SOP1_approximate_case2}
\end{equation}%
where $\mathcal{C}_{2}^{(i,k,J)}$ is given in (\ref{C2}).

\begin{itemize}
\item ${L_{R}m_{S_{i}R}}=%
{L_{E_{k}}m_{S_{J}E_{k}}}$
\end{itemize}

Replacing (\ref{M1_approximate_case2}), (\ref{M2_approximate_case2}), (\ref%
{M1_approximate_case3}), (\ref{M2_approximate_case3}), (\ref{M3_approximate}%
), and (\ref{gamma_approximate}) into (\ref{SOP_1k_final}), and by
considering $l={L_{R}m_{S_{i}R}}{\small -1,}$ $%
SOP_{1}^{(i,k,J)}$ can be approximated as 
\begin{equation}
SOP_{1}^{(i,k,J)}\sim \mathcal{C}_{3}^{(i,k,J)}\frac{\log \left( \overline{%
{\small \gamma }}_{I}\right) }{\overline{{\small \gamma }}_{I}^{%
{L_{E_{k}}m_{S_{J}E_{k}}}}},  \label{sop1_approximate_case3}
\end{equation}%
where $\mathcal{C}_{3}^{(i,k,J)}$ is given in (\ref{C3}).

\subsubsection{Case 2: Absence of a Friendly Jammer}

In order to derive the asymptotic expression of $SOP_{1}^{(i,k)}$ given in (%
\ref{SOP1_ik_final}), we need to approximate the upper incomplete Gamma
function. One can ascertain by applying the Maclaurin series that

\begin{equation}
\Gamma \left( a,b+cz\right) \sim \Gamma \left( a,b\right) -czb^{a-1}e^{-b},
\end{equation}%
as $\ z$ tends to $0.$ By considering only the two cases i.e., $l=%
{L_{R}m_{S_{i}R}}-1$ and $l={L_{R}m_{S_{i}R}}%
-2 $ and performing some algebraic manipulations, one can obtain

\begin{equation}
SOP_{1}^{(i,k)}\sim 1-\mathcal{A}_{{\small S}_{i}{\small E}_{k}{\small ,S}%
_{i}{\small R,R}}-\frac{\mathcal{A}_{{\small S}_{i}{\small E}_{k}{\small ,S}%
_{i}{\small R,S}_{i},R}}{\overline{{\small \gamma }}_{I}},
\label{SOP1_ik_asmptotic}
\end{equation}%
where $\mathcal{A}_{{\small \bullet ,\bullet ,\bullet }}$ and $\mathcal{A}_{%
{\small \bullet ,\bullet ,\bullet ,\bullet }}$ are defined in (\ref{A1}) and
(\ref{A2}), respectively.
\subsection{Asymptotic Expression of $SOP_{2}^{(k)}$}

As $SOP_{1}^{(i,k)}$ and $SOP_{2}^{(k)}$ given in (\ref{SOP1_ik_final}) and (%
\ref{SOP2_final}), respectively have the same shape, one can see that

\begin{equation}
SOP_{2}^{(k)}\sim 1-\mathcal{A}_{{\small RE}_{k}{\small ,RD}}-\frac{\mathcal{%
A}_{{\small RE}_{k}{\small ,RD,R}}}{\overline{{\small \gamma }}_{I}},
\label{SOP2_asymptotic}
\end{equation}

Finally, replacing (\ref{SOP1_approximate_case1}), (\ref%
{SOP1_approximate_case2}), (\ref{sop1_approximate_case3}), and (\ref%
{SOP2_asymptotic}) into (\ref{SOPikJ}), one can get the expressions (\ref{Asymptotic_SOPiJ_case1})-(\ref{Asymptotic_SOPiJ_case4}), respectively.
Furthermore, substituting (\ref{SOP1_ik_asmptotic}) and (\ref{SOP2_asymptotic}) into (\ref{SOP_i}), (\ref{SOP_asymptotic_final_without_jammer}) is attained which concludes the proof of Theorem 2.

\end{document}